\documentclass[10pt]{article}
\usepackage{latexsym}
\usepackage{amsfonts}
\usepackage{latexsym}
\usepackage{amsmath}
\usepackage{amssymb}
\usepackage{amsthm}

\theoremstyle{theorem}
\newtheorem{lem}{Lemma}[section]

\newtheorem{thm}[lem]{Theorem}

\theoremstyle{definition}

\begin{document}
%%%%%%%%%%%%%%%%%%%%%%%%%%%%%%%%%%%%%%%%%%%%%%%%%%%%%%%%%%%%%%%%%
\title{Symmetries of differential-difference dynamical systems in a two-dimensional lattice}

\author{Isabelle Ste-Marie\footnote{\texttt{isabelle.ste-marie@uqtr.ca}}$\ $ and S\'{e}bastien Tremblay\footnote{\texttt{sebastien.tremblay@uqtr.ca}} \\
D\'{e}partement de math\'{e}matiques et d'informatique, \\
Universit\'{e} du Qu\'{e}bec, Trois-Rivi\`{e}res, Qu\'{e}bec,\\
G9A 5H7, Canada}

\maketitle

\begin{abstract}
Classification of differential-difference equation of the form
$\ddot{u}_{nm}=F_{nm}\big(t,\ \{u_{pq}\}|_{(p,q)\in \Gamma}\big)$
are considered according to their Lie point symmetry groups. The
set $\Gamma$ represents the point $(n,m)$ and its six nearest
neighbors in a two-dimensional triangular lattice. It is shown
that the symmetry group can be at most $12$-dimensional for
abelian symmetry algebras and $13$-dimensional for nonsolvable
symmetry algebras.
\end{abstract}

\vspace{4cm} \noindent PACS numbers: 02.20.Sv, 02.30.-f, 05.45.-a, 05.50.+q \\
%\noindent Mathematics Subject Classification: 02.20.Sv, 05.45.-a, 05.50.+q\\
\noindent Keywords: Lie algebras, differential-difference
equations, two-dimensional lattice

%%%%%%%%%%%%%%%%%%%%%%%%%%%%%%%%%%%%%%%

\newpage

\section{Introduction}

The purpose of this article is to perform a symmetry analysis of the following class of differential-difference
equations

\begin{equation}
\label{equation} \Delta_{nm}\equiv \ddot{u}_{nm}-F_{nm}\big(t,\
\{u_{pq}\}|_{(p,q)\in \Gamma}\big)=0,
\end{equation}
where the overdots denote time derivative. The set $\Gamma$ being
the point $(n,m)$ and its six nearest neighbors in a
two-dimensional triangular lattice (see Figure~1) given by
$$
\label{label}
\Gamma=\Big\{(n,m),\ (n+1,m),\
(n,m+1),\ (n-1,m+1),\ (n-1,m),\ (n,m-1),\ (n+1,m-1)\Big\}.
$$

The dependent variable $u_{nm}(t)$ can be interpreted as atomic
normal displacement from their equilibrium position to the
$2$-dimensional triangular lattice, where $(n,m)$ locate the
vertex on skew coordinate system. The function $F_{nm}$, called
the \emph{interaction}, is an \emph{a priori} unspecified smooth
functions. Our aim is to classify such system according to the Lie
point symmetries that it allows, that is, to classify the
functions $F_{nm}$.

\tiny

\setlength{\unitlength}{0.5cm}
\begin{center}
\begin{picture}(20,12)
\thicklines

\put(0,0){\circle*{0.3}}

\put(3,0){\circle*{0.3}}

\put(6,0){\circle*{0.3}}

\put(9,0){\circle*{0.3}}

\put(12,0){\circle*{0.3}}

\put(1.5,2.5){\circle*{0.3}}

\put(4.5,2.5){\circle*{0.3}}
\put(3.4,1.9){(\emph{n},\emph{m}$-1$)}

\put(7.5,2.5){\circle*{0.3}}
\put(6.2,1.9){(\emph{n}$+1$,\emph{m}$-1$)}

\put(10.5,2.5){\circle*{0.3}}

\put(0,5){\circle*{0.3}}

\put(3,5){\circle*{0.3}} \put(2,4.2){(\emph{n}$-1$,\emph{m})}

\put(6,5){\circle*{0.3}} \put(5.3,4.2){(\emph{n},\emph{m})}

\put(9,5){\circle*{0.3}} \put(8,4.2){(\emph{n}$+1$,\emph{m})}

\put(12,5){\circle*{0.3}}

\put(1.5,7.5){\circle*{0.3}}

\put(4.5,7.5){\circle*{0.3}}
\put(3,6.9){(\emph{n}$-1$,\emph{m}$+1$)}

\put(7.5,7.5){\circle*{0.3}}
\put(6.5,6.9){(\emph{n},\emph{m}$+1$)}

\put(10.5,7.5){\circle*{0.3}}

\put(0,10){\circle*{0.3}}

\put(3,10){\circle*{0.3}}

\put(6,10){\circle*{0.3}}

\put(9,10){\circle*{0.3}}

\put(12,10){\circle*{0.3}}

\put(15,5){\vector(3,0){3}} \put(18.2,4.9){\emph{n}}

\put(15,5){\vector(1,1){3}} \put(18.2,8){\emph{m}}

\label{lattice}
\end{picture}
\end{center}
\normalsize

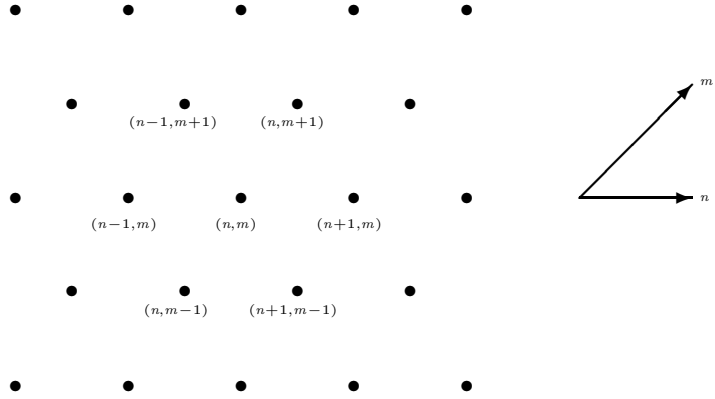
\noindent{\bf \mbox{Figure~1.}} The atoms $(n,m)$ and its six
neighbors in a two-dimensional triangular lattice.

\bigskip

\noindent The assumption for this model are the following:
\begin{enumerate}
\item[(i)] The interaction $F_{nm}$ involves only nearest
neighbors in the triangular lattice, i.e. atom $(n,m)$ interacts
only with the atoms of the set $\Gamma$, see Figure~1.

\item[(ii)] In the bulk of the article the interaction $F_{nm}$ is assumed to be nonlinear and coupled, i.e.
$$
\label{nonlin}
\frac{\partial^2F_{nm}}{\partial u_{pq}\partial u_{p'q'}}\neq 0\ \mbox{for some
$(p,q),(p',q')\in \Gamma$}\ \ \ \ \mbox{ and }\ \ \ \ \frac{\partial F_{nm}}{\partial u_{pq}}\neq 0\ \mbox{for all $(p,q)\in \Gamma$}.
$$

\item[(iii)] The interaction $F_{nm}$ is isotropic in the three
directions of the lattice.
%given by:
%\begin{enumerate}
%    \item[1)] $\triangle n\neq 0$ and $\triangle m =0$;
%    \item[2)] $\triangle n= 0$ and $\triangle m \neq 0$;
%    \item[3)] $\triangle n\neq 0$ and $\triangle m \neq 0$.
%\end{enumerate}

\item[(iv)] We suppose that interaction $F_{nm}$ depends continuously on the discrete variables $n$ and $m$. For instance, interactions depending on terms of the type $(-1)^{nm}$ are excluded.

\item[(v)] Only ``maximal'' symmetry algebras for a given
interaction $F_{nm}$ are listed. In other words, if a \emph{given}
interaction $F_{nm}$ allows symmetry algebras
$\mathcal{L}_1,\ldots,\mathcal{L}_N$ with $\dim \mathcal{L}_1<\dim
\mathcal{L}_2<\cdots <\dim \mathcal{L}_N$, then we will only list
the case $\mathcal{L}_N$.

\end{enumerate}
Our motivation is the same as for classifying differential equations according to their symmetries, see \cite{13,14,15,16}. When
(\ref{equation}) allows a nontrivial symmetry group, then it is usually possible to obtain exact analytical
solutions satisfying certain symmetry requirements. Model of this type have many applications in solid-state
physics, see for instance the work of B\"{u}ttner \emph{et al} \cite{1,2,18,19}.

\bigskip

The formalism used in this article was called ``intrinsic method''
\cite{3,4}. It has already been applied by Winternitz \emph{et al}
to the following one-dimensional cases: monoatomic molecular
chains \cite{5}, diatomic molecular chains \cite{6} and to a model
with two types of atoms distributed along a double chain \cite{7}.
In this paper we consider a two-dimensional case. In this formalism the Lie algebra of the
symmetry group, often called ``symmetry algebra'', is realized by
the following vector fields

\begin{equation}\label{vector}
\widehat
X=\tau(t,u_{nm})\partial_t+\phi_{nm}(t,u_{nm})\partial_{u_{nm}}.
\end{equation}
The algorithm for finding the functions $\tau$ and $\phi_{nm}$ is
to find the second prolongation pr$^{(2)}\widehat X$ of $\widehat
X$ and to impose that it should annihilate (\ref{equation}) on
their solution set, i.e.

\begin{equation}\label{algo}
\mathrm{pr}^{(2)}\widehat X\cdot
\Delta_{nm}\Big|_{\Delta_{nm}=0}=0.
\end{equation}

\bigskip

Our prime objective will be to find and classify all interactions $F_{nm}$ for which (\ref{equation}) allows at
least a one dimensional symmetry algebra. Thereafter, we will specify interactions further and find all those that
allow a higher dimensional symmetry algebra.

\bigskip

Note that there are other formalisms to find symmetry algebra of
differential-difference equations. Consequently the use of the
intrinsic method -- applied for fixed lattice and used here -- is
not obligatory. Another method, called the ``differential equation
approach to differential-difference equation'' \cite{4}, can also
be considered for such kind of lattices. Other methods exist in
which the group transformations can also act on the lattice
\cite{8,9,10,11}. A complete review of these different methods
have been considered in \cite{12}.

In Section 2 we present the determining equations for the
symmetries and illustrate the allowed transformation, i.e. the
classification group that will be considered to obtain class of
equations. In Section~3 all the calculations to find the
one-dimensional symmetry algebras are explicitly given. Some
notations are also introduced for the rest of the paper. Section~4
is devoted to interactions with abelian symmetry algebras. They
are denoted by $A_{jk}$, where the first index shows the dimension
of the algebra and the second index enumerates the nonequivalent
classes. Their dimensions satisfy $1 \leq \dim \mathcal{L}\leq 12$
with $\dim \mathcal{L}\neq 9,11$. Nonsolvable symmetry algebras,
denoted by $NS_{jk}$, are treated in Section~5. They all contain
sl$(2,\mathbb{R})$ as subalgebra and $3 \leq \dim \mathcal{L} \leq
13$ with $\dim \mathcal{L}\neq 10,12$. The results are summed up
in Section~6 where we also outline future work to be done.

\section{Formulation of the problem}
The second prolongation of the vector field (\ref{vector}) is
given by \cite{3,4,5}

\begin{equation}
\label{prolongation} \mathrm{pr}^{(2)}\widehat
X=\tau(t,u_{nm})\partial_t+\displaystyle \sum_{(p,q)\in \Gamma}
\phi_{pq}(t,u_{pq})\partial_{u_{pq}}+ \
\phi_{nm}^{tt}\partial_{\ddot{u}_{nm}}.
\end{equation}
The coefficient $\phi_{nm}^{tt}$ is a function depending on
$n,m,t,u_{nm},\dot{u}_{nm}$ and $\ddot{u}_{nm}$ given by

\begin{equation}
\phi_{nm}^{tt}=D_{t}^2\phi_{nm}-(D_{t}^2\tau)\,\dot{u}_{nm}-2(D_{t}\tau)\,\ddot{u}_{nm}, \label{phitt}
\end{equation}
where $D_t$ represents the total time derivative.

\bigskip

From (\ref{algo}), (\ref{prolongation}) and (\ref{phitt}) one can obtain the determining equations for the
symmetries. We eliminate the $\ddot{u}_{nm}$ terms using (\ref{equation}) and then request that the coefficients
of $\dot{u}_{nm}^3$, $\dot{u}_{nm}^2$, $\dot{u}_{nm}$ and $\dot{u}_{nm}^0$ should vanish independently. From the
determining equations of the coefficients $\dot{u}_{nm}^k$ $k=1,2,3$, the vector field (\ref{vector}) must have
the form

\begin{equation}\label{vector2}
\widehat
X=\tau(t)\partial_t+\left[\left(\frac{\dot{\tau}}{2}+a_{nm}\right)u_{nm}+\lambda_{nm}(t)\right]\partial_{u_{nm}},
\end{equation}
where $\dot{a}_{nm}=0$. The remaining determining equation
involves explicitly the interaction $F_{nm}$ and is given by

\begin{equation}\label{remaining}
\begin{array}{l}
\displaystyle
\frac{\dddot{\tau}}{2}u_{nm}+\ddot{\lambda}_{nm}+\left(a_{nm}-\frac{3}{2}\dot{\tau}\right)F_{nm}-\tau
\partial_t F_{nm} \\*[2ex]
=\ \displaystyle \sum_{(p,q)\in \Gamma}
\left[\left(\frac{\dot{\tau}}{2}+a_{pq}\right)u_{pq}+\lambda_{pq}(t)\right]\partial_{u_{pq}}F_{nm}.
\end{array}
\end{equation}
Our aim is to solve (\ref{remaining}) with respect to both the form of the nonlinear equation and the symmetry
field. For every nonlinear interaction $F_{nm}$ we wish to find the corresponding maximal symmetry group. Since
for any symmetry group there will be a whole class of nonlinear differential-difference equation related to each
other by point transformations, we will look for the simplest element of a given class. Hence, we shall classify
(\ref{equation}) into equivalence classes under the action of a group of ``allowed transformations''. We
restrict the allowed transformations to be fiber preserving, i.e. to have the form

$$
\label{allowed} \tilde{t}=\tilde{t}(t),\quad \quad
u_{nm}(t)=\Omega_{nm}\Big(t,\tilde{u}_{nm}(\tilde{t})\Big),\quad \quad (\tilde{n},\tilde{m})=(n,m),
$$
where $\Omega_{nm}$ and $\tilde{t}$ are some locally smooth and monotonous (invertible) functions. Substituting
these transformations into (\ref{equation}) and requiring that the form of the equation is preserved, we find

$$
\label{allowed2} u_{nm}(t)=\dot{\tilde t}^{-1/2}P_{nm}\ \tilde{u}_{nm}(\tilde{t})+Q_{nm}(t),
$$
where $P_{nm}\neq 0$, $\dot{P}_{nm}=0$ and $\dot{\tilde t}\neq 0$.
The transformed system is then given by
$$
\ddot{\tilde{u}}_{nm}(\tilde{t})=\tilde{F}_{nm}\big(\tilde{t},\
\{\tilde{u}_{pq}\}|_{(p,q)\in \Gamma}\big),
$$
where

$$\label{allowed3} \tilde{F}_{nm}=\dot{\tilde{t}}^{-3/2}P_{nm}^{-1}\Big(F_{nm}-\ddot{Q}_{nm}(t)\Big)+
\frac{1}{2}\dot{\tilde{t}}^{-3}\left(\dddot{\tilde{t}}-\frac{3}{2}\,\dot{\tilde{t}}^{-1}\,\ddot{\tilde{t}}^{-2}\right)
\tilde{u}_{nm}(\tilde{t}).
$$

The vector field given by (\ref{vector2}) is transformed into

$$\label{transfvect}
\begin{array}{rcl}
\widehat
X&=&\tau(t)\,\dot{\tilde{t}}\,\partial_{\tilde{t}}+\left\{\left(\frac{\dot{\tau}}{2}+
a_{nm}+\frac{1}{2}\dot{\tilde{t}}^{-1}\,\ddot{\tilde{t}}\,\,\tau\right)\tilde{u}_{nm}\right.
\\*[2ex]&+&\left.
\dot{\tilde{t}}^{1/2}P_{nm}^{-1}\left[\left(\frac{\dot{\tau}}{2}+
a_{nm}\right)Q_{nm}(t)+\lambda_{nm}(t)-\tau
\dot{Q}_{nm}\right]\right\}\partial_{\tilde{u}_{nm}}.
\end{array}
$$
In other words, under the allowed transformations the vector field
(\ref{vector2}) characterized by the triplet $\big\{\tau(t),\
a_{nm},\ \lambda_{nm}(t)\big\}$ is transformed into the triplet
$\big\{\tilde{\tau}(\tilde{t}),\ \tilde{a}_{nm},\
\tilde{\lambda}_{nm}(\tilde{t})\big\}$ where

\begin{eqnarray}
\label{transtau} \tau(t)& \rightarrow & \tilde{\tau}(\tilde{t})=\tau\big(t(\tilde{t})\big)\,\dot{\tilde{t}},
\nonumber
\\*[2ex]
\label{transa} a_{nm} & \rightarrow & \tilde{a}_{nm}=a_{nm}, \nonumber
\\*[2ex]
\label{translambda}\lambda_{nm}(t) & \rightarrow & \tilde{\lambda}_{nm}(\tilde{t})=
\dot{\tilde{t}}^{1/2}P_{nm}^{-1}\left[\left(\frac{\dot{\tau}}{2}+ a_{nm}\right)Q_{nm}(t)+\lambda_{nm}(t)-\tau
\dot{Q}_{nm}\right]. \nonumber
\end{eqnarray}

\section{One-dimensional symmetry algebras of the systems}
Let us now introduce some notations that will be useful in what follows. First we will write functions $g$ of the type $g\big(\{\xi_{pq}\}|_{(p,q)\in S}\big)$, where $S$ is a subset of $\mathbb{Z}^2$ and $\xi_{pq}:\mathbb{Z}^2\rightarrow \mathbb{R}$, more simply as $g(\xi_{pq})$ with $(p,q)\in S$. Second, if
$\{f_{pq}^{(1)},\ f_{pq}^{(2)},\ldots ,f_{pq}^{(N)}\}$ is a set of $N$ functions $f_{pq}^{(i)}: \mathbb{Z}^2\rightarrow \mathbb{R}$ depending on the discrete
variables $p,q$, then we define the determinant function
$\mathcal{D}:\mathbb{Z}^{2N}\rightarrow \mathbb{R}$ by
$$
\mathcal{D}\big[f_{p_1q_1}^{(1)},f_{p_2q_2}^{(2)},\ldots,f_{p_{N}q_{N}}^{(N)}\big]:=
\left|%
\begin{array}{cccc}
  f_{p_1q_1}^{(1)} & f_{p_2q_2}^{(1)} & \hdots & f_{p_{N}q_{N}}^{(1)} \\
  f_{p_1q_1}^{(2)} & f_{p_2q_2}^{(2)} & \hdots & f_{p_{N}q_{N}}^{(2)} \\
  \vdots & \vdots & \ddots & \vdots \\
  f_{p_1q_1}^{(N)} & f_{p_2q_2}^{(N)} & \hdots &
  f_{p_{N}q_{N}}^{(N)} \\
\end{array}%
\right|.
$$
For example, we have
$
\mathcal{D}[f_{nm},g_{n+1m},1_{nm+1}]= \left|%
\begin{array}{ccc}
  f_{nm} & f_{n+1m} & f_{nm+1} \\
  g_{nm} & g_{n+1m} & g_{nm+1} \\
  1_{nm} & 1_{n+1m} & 1_{nm+1} \\
\end{array}%
\right|$ where the function $1_{pq}:\mathbb{Z}^2\rightarrow
\mathbb{R}$ is the constant function $(p,q)\mapsto 1$. (Here, the
indices of the constant function are written only for clarity of
the determinant function $\mathcal{D}[f_{nm},g_{n+1m},1_{nm+1}]$.)

\begin{thm}
Equation $(\ref{equation})$ allows a $1$-dimensional symmetry
algebra for $3$ classes of interactions $F_{nm}$. The algebras and
interaction functions can be represented as follows:

\begin{equation}
\label{A1,1} A_{1,1}:  \quad \widehat
X=\partial_t+a_{nm}u_{nm}\partial_{u_{nm}},
\end{equation}
\begin{equation}
 \label{FA1,1}
F_{nm}=\exp(a_{nm}t)f_{nm}\big(\xi_{pq}\big),\quad \xi_{pq}=u_{pq}\exp(-a_{pq}t),\quad (p,q)\in \Gamma.
\end{equation}

\begin{equation}
\label{A1,2} A_{1,2}:  \quad \widehat
X=a_{nm}u_{nm}\partial_{u_{nm}},
\end{equation}
\begin{equation}
 \label{FA1,2}
F_{nm}=u_{nm}f_{nm}\big(t,\,\xi_{pq}\big),\quad \xi_{pq}=\displaystyle
\frac{(u_{nm})^{a_{pq}}}{(u_{pq})^{a_{nm}}},\quad (p,q)\in \Gamma\setminus\{(n,m)\}.
\end{equation}

\begin{equation}
 \label{A1,3}
 A_{1,3}:\quad
\widehat X=\lambda_{nm}(t)\,\partial_{u_{nm}},
\end{equation}
\begin{equation}
\label{FA1,3} F_{nm}=\displaystyle \frac{\ddot{\lambda}_{nm}}{\lambda_{nm}}u_{nm} +f_{nm}\big(t,\xi_{pq}\big),\,
\xi_{pq}=\mathcal{D}[u_{nm},\lambda_{pq}],\ (p,q)\in \Gamma\setminus\{(n,m)\}.
\end{equation}
\end{thm}

\noindent \textbf{Proof}. We suppose that the system
(\ref{equation}) has at least a one-dimensional symmetry group
generated by vector field of the form (\ref{vector2}). Using the
allowed transformations, we transform $\widehat X$ into three
inequivalent classes:

\begin{enumerate}
\item[a)] For the case $\tau(t)\neq 0$, we choose the functions
$\tilde{t}(t)$ and $Q_{nm}(t)$ so as to transform
$\tau(t)\rightarrow 1$ and $\lambda_{nm}(t)\rightarrow 0$. More
precisely, we look for functions $\tilde{t}(t)$ and $Q_{nm}(t)$
satisfying the following ODEs

$$
\label{taudifz}
\tau(t)\,\dot{\tilde{t}}=1, \quad \quad \tau
\dot{Q}_{nm}-\lambda_{nm}(t)-\left(\frac{\dot{\tau}}{2}+a_{nm}\right)Q_{nm}(t)=0.
$$
Hence, under allowed transformations, the vector field
(\ref{vector2}) with $\tau(t)\neq 0$ is given by (\ref{A1,1}). We
can now solve the remaining equation (\ref{remaining}), for
$\tau=1$ and $\lambda_{nm}=0$, by applying the method of
characteristics and we find function (\ref{FA1,1}).

\item[b)] For the case $\tau=0$ and $a_{nm}\neq 0$ we choose the
function $Q_{nm}(t)$ such that $\lambda_{nm}(t)\rightarrow~0$,
i.e.

$$
\label{taudifz}
a_{nm}Q_{nm}(t)+\lambda_{nm}(t)=0.
$$
The vector field (\ref{vector2}) is then given (\ref{A1,2}). The
remaining equation (\ref{remaining}) gives (\ref{FA1,2}).

\item[c)] Finally, when $\tau=0$ and $a_{nm}=0$ we already find
the vector field (\ref{A1,3}) and the remaining equation
(\ref{remaining}) gives us (\ref{FA1,3}). $\blacksquare$
\end{enumerate}

Let us notice that vector field (\ref{A1,3}) can be simplified in
some cases. Indeed, the vector field can be transformed as
$\lambda_{nm}(t)\rightarrow
\tilde{\lambda}_{nm}(\tilde{t})=\dot{\tilde{t}}^{1/2}P_{nm}^{-1}\lambda_{nm}(t)$
from the allowed transformations (\ref{translambda}). Hence, if
$\lambda_{nm}(t)$ is separable in terms of the discrete variables
$n,m$ and the continuous variable $t$, then we can transform
$\lambda_{nm}(t)$ into $1$.

We observe that the existence of a one-dimensional symmetry
algebra restricts the interaction $F_{nm}$ to arbitrary functions
of $7$~variables, rather than $8$~variables in the original
equation (\ref{equation}). In the next sections we will assume
that the interaction $F_{nm}$ and one of the symmetry generators
is already in ``canonical form'', i.e. they have the form
(\ref{A1,1}), (\ref{A1,2}) or (\ref{A1,3}) with the corresponding
interaction. We will illustrate how the interaction is further
restricted by the existence of higher dimensional symmetry
algebra.

\section{Abelian symmetry algebras}

\begin{thm}
Equation $(\ref{equation})$ allows a $2$-dimensional abelian symmetry algebra for $4$ classes of interactions
$F_{nm}$. The algebras and interaction functions can be represented as follows:
$$
A_{2,1}:\quad \widehat
X_1=\partial_t+a_{nm}^{(1)}u_{nm}\partial_{u_{nm}},\quad \widehat
X_2=a_{nm}^{(2)}u_{nm}\partial_{u_{nm}},
$$
$$
 F_{nm}=u_{nm}f_{nm}(\xi_{pq}),\quad
\xi_{pq}=\frac{(u_{pq})^{a_{nm}^{(2)}}}{(u_{nm})^{a_{pq}^{(2)}}}
\exp\big(\mathcal{D}[a_{nm}^{(1)},a_{pq}^{(2)}]t\big),\ (p,q)\in\Gamma\setminus\{(n,m)\}.
$$

$$
A_{2,2}:\quad \widehat
X_1=\partial_t+a_{nm}u_{nm}\partial_{u_{nm}},\quad \widehat
X_2=\mathrm{e}^{a_{nm}t}\partial_{u_{nm}},
$$
$$
F_{nm}=a^2_{nm}u_{nm}+\mathrm{e}^{a_{nm}t}f_{nm}(\xi_{pq}),\quad
\xi_{pq}=\mathcal{D}[u_{nm}\mathrm{e}^{-a_{nm}t},1_{pq}],\
(p,q)\in\Gamma \setminus\{(n,m)\}.
$$

$$
A_{2,3}:\quad \widehat
X_1=a^{(1)}_{nm}u_{nm}\partial_{u_{nm}},\quad \widehat
X_2=a^{(2)}_{nm}u_{nm}\partial_{u_{nm}},
$$
$$
F_{nm}=u_{nm}f_{nm}(t,\xi_{pq}),\quad (p,q)\in\Gamma \setminus\{(n,m),(n+1,m)\},
$$
$$
\xi_{pq}=\big(u_{nm}\big)^{-\mathcal{D}[a^{(1)}_{n+1m},a^{(2)}_{pq}]}\,\big(u_{n+1m}\big)^{\mathcal{D}[a^{(1)}_{nm},a^{(2)}_{pq}]}
\,\big(u_{pq}\big)^{-\mathcal{D}[a^{(1)}_{nm},a^{(2)}_{n+1m}]}.
$$

$$
A_{2,4}:\quad \widehat X_1=\partial_{u_{nm}},\quad \widehat
X_2=t\partial_{u_{nm}},
$$
$$
 F_{nm}=f_{nm}(t,\xi_{pq}),\quad
\xi_{pq}=\mathcal{D}[u_{nm},1_{pq}],\quad (p,q)\in\Gamma \setminus\{(n,m)\}. \label{FA26}
$$
\end{thm}

Two cases are not listed in these $2$-dimensional symmetry
algebras. One case corresponds to interaction $F_{nm}$ with
$\widehat X_1$ and $\widehat X_2$ of $A_{4,4}$ (see above in the
list of $4$-dimensional symmetry algebras) and is not listed here
since the symmetry algebra can be of dimension $4$ with the same
interaction, i.e. the algebra is not ``maximal''. The second case
corresponds to the degenerate case
$\mathcal{D}[\lambda_{nm}^{(1)}, \lambda_{n+1m}^{(2)}]=0$ with
$\widehat X_1$ and $\widehat X_2$ of $A_{4,4}$. The generators are
then given by $\widehat X_1=\lambda_{nm}(t)\partial_{u_{nm}}$ and
$\widehat X_2=\gamma_m(t)\,\lambda_{nm}(t)\partial_{u_{nm}}$ with
$\gamma_m\neq \gamma_{m+1}$ and $\dot \gamma_m$. Hence, we obtain
a non-isotropic system.

\begin{thm}
Equation (\ref{equation}) allows a $3$-dimensional abelian
symmetry algebra for $3$ classes of interactions $F_{nm}$. The
algebras and interaction functions can be represented as follows:
$$
A_{3,1}:\quad \widehat
X_1=\partial_t+a_{nm}^{(1)}u_{nm}\partial_{u_{nm}},\quad \widehat
X_2=a_{nm}^{(2)}u_{nm}\partial_{u_{nm}},\quad \widehat
X_3=a_{nm}^{(3)}u_{nm}\partial_{u_{nm}},
$$
$$
 F_{nm}=u_{nm}f_{nm}(\xi_{pq}),\quad
(p,q)\in\Gamma\setminus\{(n,m),(n+1,m)\},
$$
$$
\xi_{pq}=(u_{nm})^{\mathcal{D}[a_{n+1m}^{(2)},a_{pq}^{(3)}]}(u_{n+1m})^{-\mathcal{D}[a_{nm}^{(2)},a_{pq}^{(3)}]}
(u_{pq})^{\mathcal{D}[a_{nm}^{(2)},a_{n+1m}^{(3)}]}
\exp\left(-\mathcal{D}[a_{nm}^{(1)},a_{n+1m}^{(2)},a_{pq}^{(3)}]t\right).
$$

$$
A_{3,2}:\quad \widehat
X_1=\partial_t+a_{nm}u_{nm}\partial_{u_{nm}},\quad \widehat
X_2=\mathrm{e}^{a_{nm}t}\partial_{u_{nm}},\quad \widehat
X_3=\kappa_{nm}\mathrm{e}^{a_{nm}t}\partial_{u_{nm}},
$$
$$
F_{nm}=a^2_{nm}u_{nm}+\mathrm{e}^{a_{nm}t}f_{nm}(\xi_{pq}),\quad(p,q)\in\Gamma \setminus\{(n,m),(n+1,m)\}
$$
$$
\xi_{pq}=\mathcal{D}[u_{nm}\mathrm{e}^{-a_{nm}t},\kappa_{n+1m},1_{pq}],\quad
\dot{\kappa}_{nm}=0,\quad \kappa_{nm}\neq \kappa_{n+1m},\quad
\kappa_{nm}\neq \kappa_{nm+1}.
$$

$$
A_{3,3}:\quad \widehat
X_1=a_{nm}^{(1)}u_{nm}\partial_{u_{nm}},\quad \widehat
X_2=a_{nm}^{(2)}u_{nm}\partial_{u_{nm}}, \quad \widehat
X_3=a_{nm}^{(3)}u_{nm}\partial_{u_{nm}},
$$
$$
F_{nm}=u_{nm}f_{nm}(t,\xi_{pq}),\quad (p,q)\in\Gamma \setminus\{(n,m),(n+1,m),(n,m+1)\},
$$
$$
\begin{array}{ll}
\xi_{pq}=&\big(u_{nm}\big)^{\mathcal{D}[a_{n+1m}^{(1)},a_{nm+1}^{(2)},a_{pq}^{(3)}]}
\big(u_{n+1m}\big)^{-\mathcal{D}[a_{nm}^{(1)},a_{nm+1}^{(2)},a_{pq}^{(3)}]}
\big(u_{nm+1}\big)^{\mathcal{D}[a_{nm}^{(1)},a_{n+1m}^{(2)},a_{pq}^{(3)}]}\\
& \times \big(u_{pq}\big)^{-\mathcal{D}[a_{nm}^{(1)},a_{n+1m}^{(2)},a_{nm+1}^{(3)}]}.
\end{array}
$$

\end{thm}

Again here, four  $3$-dimensional Lie algebras are not listed
below. Three of them are not listed because they are not
``maximal''. The corresponding maximal algebras being given by
$A_{6,4}$, $A_{4,4}$ and $A_{4,5}$ above. The other algebra
corresponds to the degenerate case $\lambda_{n+1m}=\lambda_{nm}$
of $A_{4,5}$ (without $\widehat X_4$). The generators are then
given by $\widehat X_1=\partial_{u_{nm}}$, $\widehat
X_2=t\partial_{u_{nm}}$ and $\widehat
X_3=\lambda_{m}(t)\partial_{u_{nm}}$ with $\lambda_{m+1}\neq
\lambda_{m}$. Hence, we obtain a non-isotropic system.

The discrete function $\kappa_{nm}$ in $A_{3,2}$ depends on $n$
and $m$, otherwise we obtain a decoupled non-isotropic system.

\begin{thm}
Equation (\ref{equation}) allows a $4$-dimensional abelian
symmetry algebra for $5$ classes of interactions $F_{nm}$. The
algebras and interaction functions can be represented as follows:

$$
A_{4,1}:\quad \widehat
X_1=\partial_t+a_{nm}^{(1)}u_{nm}\partial_{u_{nm}},\quad \widehat
X_i=a_{nm}^{(i)}u_{nm}\partial_{u_{nm}},\ i=2,3,4
$$

$$
F_{nm}=u_{nm}f_{nm}(\xi_{pq}),\quad (p,q)\in\Gamma\setminus\{(n,m),(n+1,m),(n,m+1)\},
$$
$$
\begin{array}{rcl}
\xi_{pq}&=&(u_{nm})^{\mathcal{D}[a_{n+1m}^{(2)},a_{nm+1}^{(3)},a_{pq}^{(4)}]}(u_{n+1m})^{-\mathcal{D}[a_{nm}^{(2)},a_{nm+1}^{(3)},a_{pq}^{(4)}]}
(u_{nm+1})^{\mathcal{D}[a_{nm}^{(2)},a_{n+1m}^{(3)},a_{pq}^{(4)}]}\\*[2ex]
&& \times (u_{pq})^{-\mathcal{D}[a_{nm}^{(2)},a_{n+1m}^{(3)},a_{nm+1}^{(4)}]}
\exp\left(-\mathcal{D}[a_{nm}^{(1)},a_{n+1m}^{(2)},a_{nm+1}^{(3)},a_{pq}^{(4)}]t\right).
\end{array}
$$

$$
A_{4,2}:\quad \widehat
X_1=\partial_t+a_{nm}u_{nm}\partial_{u_{nm}},\quad \widehat
X_2=\mathrm{e}^{a_{nm}t}\partial_{u_{nm}},\quad \widehat
X_{i+2}=\kappa_{nm}^{(i)}\mathrm{e}^{a_{nm}t}\partial_{u_{nm}},\quad
i=1,2
$$
$$ \dot{\kappa}_{nm}^{(i)}=0,\ \kappa_{n+1m}^{(i)}\neq
\kappa_{nm}^{(i)},\ \kappa_{nm+1}^{(i)}\neq \kappa_{nm}^{(i)},
$$
$$
F_{nm}=a^2_{nm}u_{nm}+\mathrm{e}^{a_{nm}t}f_{nm}(\xi_{pq}),\quad(p,q)\in\Gamma
\setminus\{(n,m),(n+1,m),(n,m+1)\},
\\*[2ex]
$$
$$
\xi_{pq}=\mathcal{D}[u_{nm}\mathrm{e}^{-a_{nm}t},\kappa_{n+1m}^{(1)},\kappa_{nm+1}^{(2)},1_{pq}].
$$

\begin{center}$
A_{4,3}:\quad \widehat X_i=a_{nm}^{(i)}u_{nm}\partial_{u_{nm}},\
i=1,\ldots,4$
\end{center}
$$
F_{nm}=u_{nm}f_{nm}(t,\xi_{pq}),\quad (p,q)\in \{(n-1,m),(n,m-1),(n+1,m-1)\},
$$
$$
\begin{array}{rcl}
\xi_{pq}&=&\big(u_{nm}\big)^{-\mathcal{D}[a_{n+1m}^{(1)},a_{nm+1}^{(2)},a_{n-1m+1}^{(3)},a_{pq}^{(4)}]}
\big(u_{n+1m}\big)^{\mathcal{D}[a_{nm}^{(1)},a_{nm+1}^{(2)},a_{n-1m+1}^{(3)},a_{pq}^{(4)}]}
\\*[2ex]
&& \times \big(u_{nm+1}\big)^{-\mathcal{D}[a_{nm}^{(1)},a_{n+1m}^{(2)},a_{n-1m+1}^{(3)},a_{pq}^{(4)}]}
\big(u_{n-1m+1}\big)^{\mathcal{D}[a_{nm}^{(1)},a_{n+1m}^{(2)},a_{nm+1}^{(3)},a_{pq}^{(4)}]}
\\*[2ex] && \times \big(u_{pq}\big)^{-\mathcal{D}[a_{nm}^{(1)},a_{n+1m}^{(2)},a_{nm+1}^{(3)},a_{n-1m+1}^{(4)}]}.
\end{array}
$$

$$
A_{4,4}: \widehat X_i=\lambda_{nm}^{(i)}(t)\partial_{u_{nm}},\
i=1,2;\quad \widehat
Y_k=\Big(\sum_{j=1}^{2}\omega_{kj}(t)\lambda_{nm}^{(j)}(t)\Big)\partial_{u_{nm}},\
k=1,2
$$

$$
F_{nm}=1+\left|%
\begin{array}{ccc}
 u_{nm} & 1 & u_{n+1m} \\
  \lambda_{nm}^{(1)} & \ddot{\lambda}_{nm}^{(1)} & \lambda_{n+1m}^{(1)}  \\
  \lambda_{nm}^{(2)} & \ddot{\lambda}_{nm}^{(2)} & \lambda_{n+1m}^{(2)} \\
\end{array}%
\right|\left(\mathcal{D}[ \lambda_{nm}^{(1)}, \lambda_{n+1m}^{(2)}]\right)^{-1}+f_{nm}(t,\xi_{pq}),
\label{FA24}
$$
$$
\xi_{pq}=\mathcal{D}[u_{nm},\lambda_{n+1m}^{(1)},\lambda_{pq}^{(2)}],\quad (p,q)\in\Gamma
\setminus\{(n,m),(n+1,m)\},\ \mathcal{D}[ \lambda_{nm}^{(1)}, \lambda_{n+1m}^{(2)}]\neq 0. \label{InvA24}
$$
$$
\mbox{with }\displaystyle
\sum_{j=1}^2\Big(\ddot{\omega}_{kj}\lambda_{nm}^{(j)}+2\dot{\omega}_{kj}\dot{\lambda}_{nm}^{(j)}\Big)=0,\quad
\det (\dot \omega_{kj})\neq 0.
$$

%$$
%A_{4,7}: \widehat X_1=\partial_{u_{nm}},\quad \widehat X_2=t
%\partial_{u_{nm}},\quad \widehat
%X_3=\lambda_{nm}^{(1)}(t)\partial_{u_{nm}},\quad \widehat
%X_4=\lambda_{nm}^{(2)}(t)\partial_{u_{nm}},
%$$
%
%\begin{equation}
%\begin{array}{l}
%F_{nm}=\displaystyle
%\frac{\ddot{\lambda}_{nm}^{(1)}}{\lambda_{n+1m}^{(1)}-\lambda_{nm}^{(1)}}(u_{n+1m}-u_{nm})-\displaystyle
%\frac{\mathcal{D}_{1}\cdot\mathcal{D}[u_{nm},\lambda_{n+1m}^{(1)},1_{nm+1}]}{(\lambda_{n+1m}^{(1)}-
%\lambda_{nm}^{(1)})\cdot\mathcal{D}[\lambda_{nm}^{(1)},\lambda_{n+1m}^{(2)},1_{nm+1}]}\\*[2ex]\\*[2ex]
%\ \ \ \ \ \ \ \ +f_{nm}(t,\xi_{pq}),
%\end{array}
%\label{FA47}
%\end{equation}
%\begin{equation}
%\mathcal{D}_{1}=\left|%
%\begin{array}{ccc}
%\ddot{\lambda}_{nm}^{(1)}& \lambda_{nm}^{(1)}& \lambda_{n+1m}^{(1)}\\
%\ddot{\lambda}_{nm}^{(2)}& \lambda_{nm}^{(2)}& \lambda_{n+1m}^{(2)}\\
%0&1&1
%\end{array}
%\right|, \quad
%\xi_{pq}=\mathcal{D}[u_{nm},\lambda_{n+1,m}^{(1)},\lambda_{nm+1}^{(2)},1_{pq}],
%\label{InvA47}
%\end{equation}
%$$
%\quad (p,q)\in \Gamma\setminus\{(n,m),(n+1,m),(n,m+1)\}, \quad
%\lambda_{n+1m}^{(1)}\neq \lambda_{nm}^{(1)},\quad
%\mathcal{D}[\lambda_{nm}^{(1)},\lambda_{n+1m}^{(2)},1_{nm+1}]\neq
%0.
%$$

$$
A_{4,5}: \widehat X_1=\partial_{u_{nm}},\quad \widehat X_2=t
\partial_{u_{nm}},\quad \widehat
X_3=\lambda_{nm}(t)\partial_{u_{nm}},\quad \widehat
X_4=\big(\omega(t)\lambda_{nm}(t)+\sigma(t)\big)\partial_{u_{nm}},
$$
$$
F_{nm}=\frac{\ddot{\lambda}_{nm}}{\lambda_{n+1m}-\lambda_{nm}}(u_{n+1m}-u_{nm})+f_{nm}(t,\xi_{pq}),\quad \xi_{pq}=\mathcal{D}[u_{nm},\lambda_{n+1m},1_{pq}],
\label{FA36}
$$
$$
\dot{\omega}> 0, \ \ \ \ \lambda_{nm}(t)=\frac{1}{\sqrt{\dot{\omega}}}\left(c_{nm}-\frac{1}{2}\int_0^t
\frac{\ddot{\sigma}(s)}{\sqrt{\dot{\omega}(s)}}\,\mathrm{d}s\right),\quad \dot{c}_{nm}=0,
$$

$$
\lambda_{n+1m}\neq \lambda_{nm}, \nonumber
\quad (p,q)\in \Gamma\setminus\{(n,m),(n+1,m)\}.
$$
\end{thm}

\begin{thm}
Equation (\ref{equation}) allows a $5$-dimensional abelian
symmetry algebra for $3$ classes of interactions $F_{nm}$. The
algebras and interaction functions can be represented as follows:
$$
A_{5,1}:\quad \widehat X_1=
\partial_{t}+a_{nm}^{(1)}u_{nm}\partial_{u_{nm}},\quad \widehat
X_i=a_{nm}^{(i)}u_{nm}\partial_{u_{nm}},\quad i=2,\ldots,5
$$
$$
F_{nm}=u_{nm}f_{nm}(\xi_{pq}),\quad
\quad(p,q)\in\{(n-1,m),(n,m-1),(n+1,m-1)\},
$$
$$
\begin{array}{rcl}
\xi_{pq}&=&(u_{nm})^{\mathcal{D}[a_{n+1m}^{(2)},a_{nm+1}^{(3)},a_{n-1m+1}^{(4)},a_{pq}^{(5)}]}(u_{n+1m})^{-\mathcal{D}[a_{nm}^{(2)},a_{nm+1}^{(3)},a_{n-1m+1}^{(4)},a_{pq}^{(5)}]}\\*[2ex]
&& \times
(u_{nm+1})^{\mathcal{D}[a_{nm}^{(2)},a_{n+1m}^{(3)},a_{n-1m+1}^{(4)},a_{pq}^{(5)}]} (u_{n-1m+1})^{-\mathcal{D}[a_{mn}^{(2)},a_{n+1m}^{(3)},a_{nm+1}^{(4)},a_{pq}^{(5)}]}\\*[2ex]
&& \times(u_{pq})^{\mathcal{D}[a_{nm}^{(2)},a_{n+1m}^{(3)},a_{nm+1}^{(4)},a_{n-1m+1}^{(5)}]}
\exp\left(-\mathcal{D}[a_{nm}^{(1)},a_{n+1m}^{(2)},a_{nm+1}^{(3)},a_{n-1m+1}^{(4)},a_{pq}^{(5)}]t\right).
\end{array}
$$
$$
A_{5,2}:\quad \widehat
X_1=\partial_t+a_{nm}u_{nm}\partial_{u_{nm}},\quad \widehat
X_2=\mathrm{e}^{a_{nm}t}\partial_{u_{nm}},\quad \widehat
X_{i+2}=\kappa_{nm}^{(i)}\mathrm{e}^{a_{nm}t}\partial_{u_{nm}},\
i=1,2,3
$$
$$
\ \dot{\kappa}_{nm}^{(i)}=0,\ \kappa_{n+1m}^{(i)}\neq
\kappa_{nm}^{(i)},\ \kappa_{nm+1}^{(i)}\neq \kappa_{nm}^{(i)},
$$
$$
F_{nm}=a^2_{nm}u_{nm}+\mathrm{e}^{a_{nm}t}f_{nm}(\xi_{pq}),\quad(p,q)\in\{(n-1,m),(n,m-1),(n+1,m-1)\},
\\*[2ex]
$$
$$
\xi_{pq}=\mathcal{D}[u_{nm}\mathrm{e}^{-a_{nm}t},\kappa_{n+1m}^{(1)},\kappa_{nm+1}^{(2)},\kappa_{n-1m+1}^{(3)},1_{pq}].
$$

$$
A_{5,3}:\quad \widehat X_i=a_{nm}^{(i)}u_{nm}\partial_{u_{nm}},\
i=1,\ldots,5
$$
$$
F_{nm}=u_{nm}f_{nm}(t,\xi_{pq}),\quad (p,q)\in \{(n,m-1),(n+1,m-1)\},
$$
$$
\begin{array}{rcl}
\xi_{pq}&=&\big(u_{nm}\big)^{-\mathcal{D}[a_{n+1m}^{(1)},a_{nm+1}^{(2)},a_{n-1m+1}^{(3)},a_{n-1m}^{(4)},a_{pq}^{(5)}]}
\big(u_{n+1m}\big)^{\mathcal{D}[a_{nm}^{(1)},a_{n+1m}^{(2)},a_{n-1m+1}^{(3)},a_{n-1m}^{(4)},a_{pq}^{(5)}]}
\\*[2ex]
&& \times \big(u_{nm+1}\big)^{-\mathcal{D}[a_{nm}^{(1)},a_{n+1m}^{(2)},a_{n-1m+1}^{(3)},a_{n-1m}^{(4)},a_{pq}^{(5)}]}
\big(u_{n-1m+1}\big)^{\mathcal{D}[a_{nm}^{(1)},a_{n+1m}^{(2)},a_{nm+1}^{(3)},a_{n-1m}^{(4)},a_{pq}^{(5)}]}
\\*[2ex]
&& \times \big(u_{n-1m}\big)^{-\mathcal{D}[a_{nm}^{(1)},a_{n+1m}^{(2)},a_{nm+1}^{(3)},a_{n-1m+1}^{(4)},a_{pq}^{(5)}]} \big(u_{pq}\big)^{\mathcal{D}[a_{nm}^{(1)},a_{n+1m}^{(2)},a_{nm+1}^{(3)},a_{n-1m+1}^{(4)},a_{n-1m}^{(5)}]}.
\end{array}
$$

\end{thm}
$$
$$

\begin{thm}
Equation (\ref{equation}) allows a $6$-dimensional abelian
symmetry algebra for $5$ classes of interactions $F_{nm}$. The
algebras and interaction functions can be represented as follows:

$$
A_{6,1}:\quad \widehat X_1=
\partial_{t}+a_{nm}^{(1)}u_{nm}\partial_{u_{nm}}, \quad \widehat
X_i=a_{nm}^{(i)}u_{nm}\partial_{u_{nm}},\quad i=2,\ldots,6
$$
$$
F_{nm}=u_{nm}f_{nm}(\xi_{pq}),\quad (p,q)\in\{(n,m-1),(n+1,m-1)\}
$$
$$
\begin{array}{rcl}
\xi_{pq}&=&(u_{nm})^{\mathcal{D}[a_{n+1m}^{(2)},a_{nm+1}^{(3)},a_{n-1m+1}^{(4)},a_{n-1m}^{(5)},a_{pq}^{(6)}]}(u_{n+1m})^{-\mathcal{D}[a_{nm}^{(2)},a_{nm+1}^{(3)},a_{n-1m+1}^{(4)},a_{n-1m}^{(5)},a_{pq}^{(6)}]}\\*[2ex]
&& \times
(u_{nm+1})^{\mathcal{D}[a_{nm}^{(2)},a_{n+1m}^{(3)},a_{n-1m+1}^{(4)},a_{n-1m}^{(5)},a_{pq}^{(6)}]} (u_{n-1m+1})^{-\mathcal{D}[a_{mn}^{(2)},a_{n+1m}^{(3)},a_{nm+1}^{(4)},a_{n-1m}^{(5)},a_{pq}^{(6)}]}\\*[2ex]
&& \times(u_{n-1m})^{\mathcal{D}[a_{nm}^{(2)},a_{n+1m}^{(3)},a_{nm+1}^{(4)},a_{n-1m+1}^{(5)},a_{pq}^{(6)}]} (u_{pq})^{-\mathcal{D}[a_{nm}^{(2)},a_{n+1m}^{(3)},a_{nm+1}^{(4)},a_{n-1m+1}^{(5)},a_{n-1m}^{(6)}]}\\*[2ex]
&& \times \exp\left(-\mathcal{D}[a_{nm}^{(1)},a_{n+1m}^{(2)},a_{nm+1}^{(3)},a_{n-1m+1}^{(4)},a_{n-1m}^{(5)},a_{pq}^{(6)}]t\right).
\end{array}
$$

$$
A_{6,2}:\quad \widehat
X_1=\partial_t+a_{nm}u_{nm}\partial_{u_{nm}},\quad \widehat
X_2=\mathrm{e}^{a_{nm}t}\partial_{u_{nm}},\quad \widehat
X_{i+2}=\kappa_{nm}^{(i)}\mathrm{e}^{a_{nm}t}\partial_{u_{nm}},\
i=1,\ldots,4
$$
$$
\ \dot{\kappa}_{nm}^{(i)}=0,\ \kappa_{n+1m}^{(i)}\neq
\kappa_{nm}^{(i)},\ \kappa_{nm+1}^{(i)}\neq \kappa_{nm}^{(i)},
$$
$$
F_{nm}=a^2_{nm}u_{nm}+\mathrm{e}^{a_{nm}t}f_{nm}(\xi_{pq}),\quad(p,q)\in\{(n,m-1),(n+1,m-1)\},
$$

$$
\xi_{pq}=\mathcal{D}[u_{nm}\mathrm{e}^{-a_{nm}t},\kappa_{n+1m}^{(1)},\kappa_{nm+1}^{(2)},\kappa_{n-1m+1}^{(3)},\kappa_{n-1m}^{(4)},1_{pq} ].
$$

$$
A_{6,3}:\quad \widehat
X_i=a_{nm}^{(i)}u_{nm}\partial_{u_{nm}},\quad i=1,\ldots,6; \quad
F_{nm}=u_{nm}f_{nm}(t,\xi),
$$
$$
\xi=\big(u_{nm}\big)^{-\mathcal{D}[a_{n+1m}^{(1)},a_{nm+1}^{(2)},a_{n-1m+1}^{(3)},a_{n-1m}^{(4)},a_{nm-1}^{(5)},a_{n+1,m-1}^{(6)}]}
\big(u_{n+1m}\big)^{\mathcal{D}[a_{nm}^{(1)},a_{nm+1}^{(2)},a_{n-1m+1}^{(3)},a_{n-1m}^{(4)},a_{nm-1}^{(5)},a_{n+1m-1}^{(6)}]}
$$
$$
\quad \times \big(u_{nm+1}\big)^{-\mathcal{D}[a_{nm}^{(1)},a_{n+1m}^{(2)},a_{n-1m+1}^{(3)},a_{n-1m}^{(4)},a_{nm-1}^{(5)},a_{n+1m-1}^{(6)}]}
\big(u_{n-1m+1}\big)^{\mathcal{D}[a_{nm}^{(1)},a_{n+1m}^{(2)},a_{nm+1}^{(3)},a_{n-1m}^{(4)},a_{nm-1}^{(5)},a_{n+1m-1}^{(6)}]}
$$
$$
\quad \times
\big(u_{n-1m}\big)^{-\mathcal{D}[a_{nm}^{(1)},a_{n+1m}^{(2)},a_{nm+1}^{(3)},a_{n-1m+1}^{(4)},a_{nm-1}^{(5)}, a_{n+1m-1}^{(6)}]}
\big(u_{nm-1}\big)^{\mathcal{D}[a_{nm}^{(1)},a_{n+1m}^{(2)},a_{nm+1}^{(3)},a_{n-1m+1}^{(4)},a_{n-1m}^{(5)}, a_{n+1m-1}^{(6)}]}
$$
$$
\quad \times  \big(u_{n+1m-1}\big)^{-\mathcal{D}[a_{nm}^{(1)},a_{n+1m}^{(2)},a_{nm+1}^{(3)},a_{n-1m+1}^{(4)},a_{n-1m}^{(5)},a_{nm-1}^{(6)}]}.
$$

$$
A_{6,4}:\quad \widehat
X_i=\lambda_{nm}^{(i)}(t)\partial_{u_{nm}},\ i=1,2,3; \quad
\widehat Y_k=\Big(\displaystyle \sum_{j=1}^3
\omega_{kj}(t)\lambda_{nm}^{(j)}(t)\Big)\partial_{u_{nm}},\
k=1,2,3
$$
$$
F_{nm}=1+\left|%
\begin{array}{cccc}
  u_{nm} & 1 & u_{n+1m} & u_{nm+1} \\
  \lambda_{nm}^{(1)} & \ddot{\lambda}_{nm}^{(1)} & \lambda_{n+1m}^{(1)} &  \lambda_{nm+1}^{(1)}\\
  \lambda_{nm}^{(2)} & \ddot{\lambda}_{nm}^{(2)} & \lambda_{n+1m}^{(2)} &  \lambda_{nm+1}^{(2)}\\
\lambda_{nm}^{(3)} & \ddot{\lambda}_{nm}^{(3)} & \lambda_{n+1m}^{(3)} & \lambda_{nm+1}^{(3)}
\end{array}
\right| \left(\mathcal{D}[\lambda_{nm}^{(1)},\lambda_{n+1m}^{(2)},\lambda_{nm+1}^{(3)}]\right)^{-1}+f_{nm}(t,\xi_{pq})
 \label{FA34}
$$
$$
\begin{array}{c}
\xi_{pq}=\mathcal{D}[u_{nm},\lambda_{n+1m}^{(1)},\lambda_{nm+1}^{(2)},\lambda_{pq}^{(3)}],\
(p,q)\in\Gamma \setminus\{(n,m),(n+1,m),(n,m+1)\}, \\*[2ex]
\mathcal{D}[\lambda_{nm}^{(1)},\lambda_{n+1m}^{(2)},\lambda_{nm+1}^{(3)}]\neq
0,\ \mathcal{D}[\lambda_{nm}^{(1)},\lambda_{n+1m}^{(2)}]\neq
0,\\*[2ex] \mbox{with }\displaystyle
\sum_{j=1}^3\Big(\ddot{\omega}_{kj}\lambda_{nm}^{(j)}+2\dot{\omega}_{kj}\dot{\lambda}_{nm}^{(j)}
\Big)=0,\quad \det (\dot \omega_{kj})\neq 0.
 \label{InvA34}
\end{array}
$$

%$$
%A_{6,8}:\quad \widehat
%X_i=\lambda_{nm}^{(i)}(t)\partial_{u_{nm}},\ i=1,2,\quad \widehat
%X_3=\Big(\displaystyle \sum_{j=1}^2
%\omega_j(t)\lambda_{nm}^{(j)}(t)\Big)\partial_{u_{nm}},\ \widehat
%X_4=\Big(\displaystyle \sum_{j=1}^2
%\omega_{j+2}(t)\lambda_{nm}^{(j)}(t)\Big)\partial_{u_{nm}},
%$$
%$$\widehat
%X_5=\Big(\displaystyle \sum_{j=1}^2
%\omega_{j+4}(t)\lambda_{nm}^{(j)}(t)\Big)\partial_{u_{nm}},\quad
%\widehat X_6=\Big(\displaystyle \sum_{j=1}^2
%\omega_{j+6}(t)\lambda_{nm}^{(j)}(t)\Big)\partial_{u_{nm}},
%$$
%$$
%\mbox{with }\displaystyle
%\sum_{j=1}^2\Big(\ddot{\omega}_{j+k}\lambda_{nm}^{(j)}+2\dot{\omega}_{j+k}\dot{\lambda}_{nm}^{(j)}
%\Big)=0 \mbox{ for } k=0,2,4,6
%$$
%$$
%F_{nm} \mbox{ as in (\ref{FA24}) and (\ref{InvA24}),with same conditions.}
%$$

%$$
%A_{6,10}: \widehat X_1=\partial_{u_{nm}},\quad \widehat X_2=
%t\partial_{u_{nm}}, \quad \widehat
%X_{i+2}=\lambda_{nm}^{(i)}(t)\partial_{u_{nm}},\ i=1,2,3
%$$
%$$
%\widehat X_{6}=\Big(\displaystyle \sum_{j=1}^3
%\big(\omega_{j}(t)\lambda_{nm}^{(j)}(t)\big)+\omega_4(t)\Big)\partial_{u_{nm}},
%$$
%$$
%\mbox{with }\displaystyle
%\sum_{j=1}^3\Big(\ddot{\omega}_j\lambda_{nm}^{(j)}+2\dot{\omega}_j\dot{\lambda}_{nm}^{(j)}
%\Big)+\ddot{\omega}_4=0,
%$$
%$$
%F_{nm} \mbox{ as in (\ref{FA58}) and (\ref{InvA58}),with same conditions.}
%$$

$$
A_{6,5}: \widehat X_1=\partial_{u_{nm}},\quad \widehat X_2=t
\partial_{u_{nm}},\quad \widehat
X_{i+2}=\lambda_{nm}^{(i)}(t)\partial_{u_{nm}},\quad i=1,2
$$
$$
\widehat Y_k=\Big(\displaystyle \sum_{j=1}^2
\omega_{kj}(t)\lambda_{nm}^{(j)}(t)+\sigma^{(k)}(t)\Big)\partial_{u_{nm}},\ k=1,2
$$
$$
\begin{array}{l}
F_{nm}=\displaystyle
\frac{\ddot{\lambda}_{nm}^{(1)}}{\lambda_{n+1m}^{(1)}-\lambda_{nm}^{(1)}}(u_{n+1m}-u_{nm})-\displaystyle
\frac{\mathcal{D}_{1}\cdot\mathcal{D}[u_{nm},\lambda_{n+1m}^{(1)},1_{nm+1}]}{(\lambda_{n+1m}^{(1)}-
\lambda_{nm}^{(1)})\cdot\mathcal{D}[\lambda_{nm}^{(1)},\lambda_{n+1m}^{(2)},1_{nm+1}]}\\*[2ex]\\*[2ex]
\ \ \ \ \ \ \ \ +f_{nm}(t,\xi_{pq}),
\end{array}
\label{FA47}
$$
$$
\mathcal{D}_{1}=\left|%
\begin{array}{ccc}
\ddot{\lambda}_{nm}^{(1)}& \lambda_{nm}^{(1)}& \lambda_{n+1m}^{(1)}\\
\ddot{\lambda}_{nm}^{(2)}& \lambda_{nm}^{(2)}& \lambda_{n+1m}^{(2)}\\
0&1&1
\end{array}
\right|, \quad
\xi_{pq}=\mathcal{D}[u_{nm},\lambda_{n+1,m}^{(1)},\lambda_{nm+1}^{(2)},1_{pq}],
\label{InvA47}
$$
$$
\quad (p,q)\in \Gamma\setminus\{(n,m),(n+1,m),(n,m+1)\}, \quad
\lambda_{n+1m}^{(1)}\neq \lambda_{nm}^{(1)},\quad
\mathcal{D}[\lambda_{nm}^{(1)},\lambda_{n+1m}^{(2)},1_{nm+1}]\neq
0.
$$

$$
\mbox{with }\displaystyle
\sum_{j=1}^2\Big(\ddot{\omega}_{kj}\lambda_{nm}^{(j)}+2\dot{\omega}_{kj}\dot{\lambda}_{nm}^{(j)}
\Big)+\ddot{\sigma}^{(k)}=0,\quad \det (\dot \omega_{kj})\neq 0.
$$

%$$
%A_{6,12}: \widehat X_1=\partial_{u_{nm}},\quad \widehat X_2=t
%\partial_{u_{nm}}, \quad \widehat
%X_3=\lambda_{nm}(t)\partial_{u_{nm}},\quad \widehat
%X_4=\Big(\omega_1(t)\lambda_{nm}(t)+\omega_2(t)\Big)\partial_{u_{nm}}
%$$
%$$
%\widehat
%X_5=\Big(\omega_3(t)\lambda_{nm}(t)+\omega_4(t)\Big)\partial_{u_{nm}},\quad
%\widehat
%X_6=\Big(\omega_5(t)\lambda_{nm}(t)+\omega_6(t)\Big)\partial_{u_{nm}},
%$$
%$$
%\mbox{with }\
%\ddot{\omega}_{1+k}\lambda_{nm}+2\dot{\omega}_{1+k}\dot{\lambda}_{nm}+\ddot{\omega}_{2+k}=0
%\mbox{ for }k=0,2,4
%$$
%$$
%F_{nm} \mbox{ as in (\ref{FA36}) and (\ref{InvA36}),with same conditions.}
%$$

\end{thm}

\begin{thm}
Equation (\ref{equation}) allows a $7$-dimensional abelian
symmetry algebra for $2$ classes of interactions $F_{nm}$. The
algebras and interaction functions can be represented as follows:

$$
A_{7,1}:\quad \widehat X_1=
\partial_{t}+a_{nm}^{(1)}u_{nm}\partial_{u_{nm}},\quad \widehat
X_i=a_{nm}^{(i)}u_{nm}\partial_{u_{nm}},\quad i=2,\ldots,7
$$
$$
F_{nm}=u_{nm}f_{nm}(\xi),
$$
$$
\begin{array}{l}
\xi=(u_{nm})^{-\mathcal{D}[a_{n+1m}^{(2)},a_{nm+1}^{(3)},a_{n-1m+1}^{(4)},a_{n-1m}^{(5)},a_{nm-1}^{(6)},a_{n+1,m-1}^{(7)}]}(u_{n+1m})^{\mathcal{D}[a_{nm}^{(2)},a_{nm+1}^{(3)},a_{n-1m+1}^{(4)},a_{n-1m}^{(5)},a_{nm-1}^{(6)},a_{n+1,m-1}^{(7)}]}\\*[2ex]
\times
(u_{nm+1})^{-\mathcal{D}[a_{nm}^{(2)},a_{n+1m}^{(3)},a_{n-1m+1}^{(4)},a_{n-1m}^{(5)},a_{nm-1}^{(6)},a_{n+1,m-1}^{(7)}]}
(u_{n-1m+1})^{\mathcal{D}[a_{mn}^{(2)},a_{n+1m}^{(3)},a_{nm+1}^{(4)},a_{n-1m}^{(5)},a_{nm-1}^{(6)},a_{n+1,m-1}^{(7)}]}\\*[2ex]
\times(u_{n-1m})^{-\mathcal{D}[a_{nm}^{(2)},a_{n+1m}^{(3)},a_{nm+1}^{(4)},a_{n-1m+1}^{(5)},a_{nm-1}^{(6)},a_{n+1,m-1}^{(7)}]}
(u_{nm-1})^{\mathcal{D}[a_{nm}^{(2)},a_{n+1m}^{(3)},a_{nm+1}^{(4)},a_{n-1m+1}^{(5)},a_{n-1m}^{(6)}a_{n+1,m-1}^{(7)}]}\\*[2ex]
\times(u_{n+1,m-1})^{-\mathcal{D}[a_{nm}^{(2)},a_{n+1m}^{(3)},a_{nm+1}^{(4)},a_{n-1m+1}^{(5)},a_{n-1m}^{(6)},a_{nm-1}^{(7)}]}\\*[2ex]
\times
\exp\left(-\mathcal{D}[a_{nm}^{(1)},a_{n+1m}^{(2)},a_{nm+1}^{(3)},a_{n-1m+1}^{(4)},a_{n-1m}^{(5)},a_{nm-1}^{(6)},a_{n+1,m-1}^{(7)}]t\right).
\end{array}
$$

$$
A_{7,2}:\quad \widehat
X_1=\partial_t+a_{nm}u_{nm}\partial_{u_{nm}},\quad \widehat
X_2=\mathrm{e}^{a_{nm}t}\partial_{u_{nm}},\quad \widehat
X_{i+2}=\kappa_{nm}^{(i)}\mathrm{e}^{a_{nm}t}\partial_{u_{nm}},\
i=1,\ldots,5
$$
$$
F_{nm}=a^2_{nm}u_{nm}+\mathrm{e}^{a_{nm}t}f_{nm}(\xi),\
\dot{\kappa}_{nm}^{(i)}=0,\ \kappa_{n+1m}^{(i)}\neq
\kappa_{nm}^{(i)},\ \kappa_{nm+1}^{(i)}\neq \kappa_{nm}^{(i)},
$$
$$
\xi=\mathcal{D}[u_{nm}\mathrm{e}^{-a_{nm}t},\kappa_{n+1m}^{(1)},\kappa_{nm+1}^{(2)},\kappa_{n-1m+1}^{(3)},\kappa_{n-1m}^{(4)},\kappa_{nm-1}^{(5)},1_{n+1m-1}
].
$$

\end{thm}

\begin{thm}
Equation (\ref{equation}) allows a $8$-dimensional abelian
symmetry algebra for $2$ classes of interactions $F_{nm}$. The
algebras and interaction functions can be represented as follows:

$$
A_{8,1}:\ \widehat X_i=\lambda_{nm}^{(i)}(t)\partial_{u_{nm}},\
i=1,\ldots,4; \quad \widehat Y_k=\Big(\displaystyle \sum_{j=1}^4
\omega_{kj}(t)\lambda_{nm}^{(j)}(t)\Big)\partial_{u_{nm}},\
k=1,\ldots,4
$$
$$
\begin{array}{l}
F_{nm}=1+\left|%
\begin{array}{ccccc}
  u_{nm} & 1 & u_{n+1m} & u_{nm+1} & u_{n-1m+1}\\
  \lambda_{nm}^{(1)} & \ddot{\lambda}_{nm}^{(1)} & \lambda_{n+1m}^{(1)} &  \lambda_{nm+1}^{(1)} &  \lambda_{n-1m+1}^{(1)}\\
  \lambda_{nm}^{(2)} & \ddot{\lambda}_{nm}^{(2)} & \lambda_{n+1m}^{(2)} &  \lambda_{nm+1}^{(2)} &  \lambda_{n-1m+1}^{(2)}\\
\lambda_{nm}^{(3)} & \ddot{\lambda}_{nm}^{(3)} & \lambda_{n+1m}^{(3)} & \lambda_{nm+1}^{(3)} & \lambda_{n-1m+1}^{(3)} \\
\lambda_{nm}^{(4)} & \ddot{\lambda}_{nm}^{(4)} & \lambda_{n+1m}^{(4)} &  \lambda_{nm+1}^{(4)} &
\lambda_{n-1m+1}^{(4)}
\end{array}
\right|\left(\mathcal{D}[\lambda_{nm}^{(1)},\lambda_{n+1m}^{(2)},\lambda_{nm+1}^{(3)},\lambda_{n-1m+1}^{(4)}]\right)^{-1}\\*[2ex]\\
\quad \quad \quad +\ f_{nm}(t,\xi_{pq}),\quad (p,q)\in \{(n-1,m),(n,m-1),(n+1,m-1)\},
\end{array}
\label{FA44}
$$
$$
\begin{array}{c}
\xi_{pq}=\mathcal{D}[u_{nm},\lambda_{n+1m}^{(1)},\lambda_{nm+1}^{(2)},\lambda_{n-1m+1}^{(3)},\lambda_{pq}^{(4)}],\quad
\mathcal{D}[\lambda_{nm}^{(1)},\lambda_{n+1m}^{(2)},\lambda_{nm+1}^{(3)},\lambda_{n-1m+1}^{(4)}]\neq
0,\\*[2ex] \ \ \ \
\mathcal{D}[\lambda_{nm}^{(1)},\lambda_{n+1m}^{(2)},\lambda_{nm+1}^{(3)}]\neq
0,\quad \mathcal{D}[\lambda_{nm}^{(1)},\lambda_{n+1m}^{(2)}]\neq
0,\\ \mbox{with }\displaystyle
\sum_{j=1}^4\Big(\ddot{\omega}_{kj}\lambda_{nm}^{(j)}+2\dot{\omega}_{kj}\dot{\lambda}_{nm}^{(j)}
\Big)=0,\quad \det (\dot \omega_{kj})\neq 0.
\end{array}
\label{InvA44}
$$

$$
A_{8,2}: \widehat X_1=\partial_{u_{nm}},\quad \widehat X_2=t
\partial_{u_{nm}}, \quad \widehat
X_{i+2}=\lambda_{nm}^{(i)}(t)\partial_{u_{nm}},\ i=1,2,3
$$
$$
\quad \widehat Y_k=\Big(\displaystyle \sum_{j=1}^3 \omega_{kj}(t)
\lambda_{nm}^{(j)}(t)+\sigma^{(k)}(t)\Big)\partial_{u_{nm}},\
k=1,2,3
$$
$$
\begin{array}{rcl}
F_{nm}&=&  \displaystyle
\frac{\ddot{\lambda}_{nm}^{(1)}}{\lambda_{n+1m}^{(1)}-\lambda_{nm}^{(1)}}(u_{n+1m}-u_{nm})+
\displaystyle
\frac{\mathcal{D}_{2}\cdot\mathcal{D}[u,\lambda_{n+1m}^{(1)},\lambda_{nm+1}^{(2)},1_{n-1m+1}]}
{(\lambda_{n+1m}^{(1)}-\lambda_{nm}^{(1)})\cdot\mathcal{D}[\lambda_{nm}^{(1)},\lambda_{n+1m}^{(2)},\lambda_{nm+1}^{(3)},1_{n-1m+1}]}\\*[2ex]
\\*[2ex] && - \displaystyle
\frac{\mathcal{D}_{3}\cdot\mathcal{D}[\lambda_{nm}^{(1)},\lambda_{n+1m}^{(3)},1_{nm+1}]
\cdot\mathcal{D}[u,\lambda_{n+1m}^{(1)},\lambda_{nm+1}^{(2)},1_{n-1m+1}]}
{(\lambda_{n+1m}^{(1)}-\lambda_{nm}^{(1)})\cdot\mathcal{D}[\lambda_{nm}^{(1)},\lambda_{n+1m}^{(2)},1_{nm+1}]\cdot\mathcal{D}[\lambda_{nm}^{(1)},
\lambda_{n+1m}^{(2)},\lambda_{nm+1}^{(3)},1_{n-1m+1}]}\\*[2ex]\\*[2ex]
&& -\displaystyle
\frac{\mathcal{D}_{3}\cdot\mathcal{D}[u_{nm},\lambda_{n+1m}^{(1)},1_{nm+1}]}{(\lambda_{n+1m}^{(1)}-
\lambda_{nm}^{(1)})\cdot\mathcal{D}[\lambda_{nm}^{(1)},\lambda_{n+1m}^{(2)},1_{nm+1}]}+
f_{nm}(t,\xi_{pq}),\\*[2ex]
\end {array}
\label{FA58}
$$
$$
\xi_{pq}=
\mathcal{D}[u_{nm},\lambda_{n+1m}^{(1)},\lambda_{nm+1}^{(2)},\lambda_{n-1m+1}^{(3)},1_{pq}]
\label{InvA58},\quad (p,q)\in \{(n-1,m),(n,m-1),(n+1,m-1)\},
$$

$$
\mathcal {D}_{2}=\left|%
\begin{array}{ccc}
\ddot{\lambda}_{nm}^{(1)}& \lambda_{nm}^{(1)}& \lambda_{n+1m}^{(1)}\\
\ddot{\lambda}_{nm}^{(2)}& \lambda_{nm}^{(2)}& \lambda_{n+1m}^{(2)}\\
0&1&1
\end{array}
\right|,\quad \mathcal {D}_{3}=\left|%
\begin{array}{ccc}
\ddot{\lambda}_{nm}^{(1)}& \lambda_{nm}^{(1)}& \lambda_{n+1m}^{(1)}\\
\ddot{\lambda}_{nm}^{(3)}& \lambda_{nm}^{(3)}& \lambda_{n+1m}^{(3)}\\
0&1&1
\end{array}
\right|,
$$
$$
\lambda_{n+1m}^{(1)} \neq \lambda_{nm}^{(1)},\quad
\mathcal{D}[\lambda_{nm}^{(1)},\lambda_{n+1m}^{(2)},1_{nm+1}]\neq
0 ,\quad
\mathcal{D}[\lambda_{nm}^{(1)},\lambda_{n+1m}^{(2)},\lambda_{nm+1}^{(3)},1_{n-1m+1}]\neq
0.
$$
$$
\mbox{with }\displaystyle
\sum_{j=1}^3\Big(\ddot{\omega}_{kj}\lambda_{nm}^{(j)}+2\dot{\omega}_{kj}\dot{\lambda}_{nm}^{(j)}
\Big)+\ddot{\sigma}^{(k)}=0,\quad \det (\dot \omega_{kj})\neq 0.
$$
\end{thm}

\begin{thm}
\label{A10}
Equation (\ref{equation}) allows a $10$-dimensional abelian
symmetry algebra for $2$ classes of interactions $F_{nm}$. The
algebras and interaction functions can be represented as follows:
\begin{equation}
\label{alg10}
A_{10,1}:\ \widehat X_i=\lambda_{nm}^{(i)}(t)\partial_{u_{nm}},\
i=1,\ldots,5; \quad \widehat Y_k=\Big(\displaystyle \sum_{j=1}^5
\omega_{kj}(t)\lambda_{nm}^{(j)}(t)\Big)\partial_{u_{nm}},\ k=1,\ldots,5
\end{equation}
\begin{equation}
\begin{array}{l}
F_{nm}=\tiny 1+\left|%
\begin{array}{cccccc}
  u_{nm} & 1 & u_{n+1m} & u_{nm+1} & u_{n-1m+1}& u_{n-1m}\\
  \lambda_{nm}^{(1)} & \ddot{\lambda}_{nm}^{(1)} & \lambda_{n+1m}^{(1)} &  \lambda_{nm+1}^{(1)} &  \lambda_{n-1m+1}^{(1)} & \lambda_{n-1m}^{(1)} \\
  \lambda_{nm}^{(2)} & \ddot{\lambda}_{nm}^{(2)} & \lambda_{n+1m}^{(2)} &  \lambda_{nm+1}^{(2)} &  \lambda_{n-1m+1}^{(2)}& \lambda_{n-1m}^{(2)}\\
\lambda_{nm}^{(3)} & \ddot{\lambda}_{nm}^{(3)} & \lambda_{n+1m}^{(3)} & \lambda_{nm+1}^{(3)} & \lambda_{n-1m+1}^{(3)} &\lambda_{n-1m}^{(3)}\\
\lambda_{nm}^{(4)} & \ddot{\lambda}_{nm}^{(4)} & \lambda_{n+1m}^{(4)} &  \lambda_{nm+1}^{(4)} & \lambda_{n-1m+1}^{(4)} & \lambda_{n-1m}^{(4)}\\
\lambda_{nm}^{(5)} & \ddot{\lambda}_{nm}^{(5)} & \lambda_{n+1m}^{(5)} &  \lambda_{nm+1}^{(5)} & \lambda_{n-1m+1}^{(5)} & \lambda_{n-1m}^{(5)}
\end{array}%
\right|
{\left(\mathcal{D}[\lambda_{nm}^{(1)},\lambda_{n+1m}^{(2)},\lambda_{nm+1}^{(3)},\lambda_{n-1m+1}^{(4)},\lambda_{n-1m}^{(5)}]\right)^{-1}}\\*[2ex]\\
\quad \quad \quad \normalsize +\ f_{nm}(t,\xi_{pq}),
\end{array}
\label{FA54}
\end{equation}
$$
\label{InvA54}
\begin{array}{c}
\xi_{pq}=\mathcal{D}[u_{nm},\lambda_{n+1m}^{(1)},\lambda_{nm+1}^{(2)},\lambda_{n-1m+1}^{(3)},\lambda_{n-1m}^{(4)},\lambda_{pq}^{(5)}], \quad (p,q)\in \{(n,m-1),(n+1,m-1)\},
\end{array}
$$
$$
\begin{array}{c}
\mathcal{D}[\lambda_{nm}^{(1)},\lambda_{n+1m}^{(2)},\lambda_{nm+1}^{(3)},\lambda_{n-1m+1}^{(4)},\lambda_{n-1m}^{(5)}]\neq 0, \quad
\mathcal{D}[\lambda_{nm}^{(1)},\lambda_{n+1m}^{(2)},\lambda_{nm+1}^{(3)},\lambda_{n-1m+1}^{(4)}]\neq 0,
\\*[2ex]  \mathcal{D}[\lambda_{nm}^{(1)},\lambda_{n+1m}^{(2)},\lambda_{nm+1}^{(3)}]\neq 0,\quad
\mathcal{D}[\lambda_{nm}^{(1)},\lambda_{n+1m}^{(2)}]\neq 0,
\end{array}
$$
\begin{equation}
\label{cond101}
\mbox{with }\displaystyle
\sum_{j=1}^5\Big(\ddot{\omega}_{kj}\lambda_{nm}^{(j)}+2\dot{\omega}_{kj}\dot{\lambda}_{nm}^{(j)}
\Big)=0,\ \det (\dot \omega_{kj})\neq 0.
\end{equation}

$$
A_{10,2}: \widehat X_1=\partial_{u_{nm}},\quad \widehat X_2=t
\partial_{u_{nm}}, \quad \widehat
X_{i+2}=\lambda_{nm}^{(i)}(t)\partial_{u_{nm}},\ i=1,\ldots,4
$$
$$
\quad \widehat Y_k=\Big(\displaystyle
\sum_{j=1}^4\big(\omega_{kj}(t)
\lambda_{nm}^{(j)}(t)\big)+\sigma^{(k)}(t)\Big)\partial_{u_{nm}},\
k=1,\ldots,4
$$
$$
\begin{array}{l}
F_{nm}= \displaystyle
\frac{\ddot{\lambda}_{nm}^{(1)}}{\lambda_{n+1m}^{(1)}-\lambda_{nm}^{(1)}}(u_{n+1m}-u_{nm})-\displaystyle
\frac{\mathcal{D}_{3}\cdot\mathcal{D}[u_{nm},\lambda_{n+1m}^{(1)},1_{nm+1}]}{(\lambda_{n+1m}^{(1)}-
\lambda_{nm}^{(1)})\cdot\mathcal{D}[\lambda_{nm}^{(1)},\lambda_{n+1m}^{(2)},1]}\\*[2ex]\\*[2ex]
+ \displaystyle \left(\mathcal{D}_{2} - \displaystyle
\frac{\mathcal{D}_{3}\cdot\mathcal{D}[\lambda_{nm}^{(1)},\lambda_{n+1m}^{(3)},1_{nm+1}]}
{\mathcal{D}[\lambda_{nm}^{(1)},\lambda_{n+1m}^{(2)},1_{nm+1}]}\right)
\displaystyle \left(
\frac{\mathcal{D}[u_{nm},\lambda_{n+1m}^{(1)},\lambda_{nm+1}^{(2)},1_{n-1m+1}]}
{(\lambda_{n+1m}^{(1)}-\lambda_{nm}^{(1)})\cdot\mathcal{D}[\lambda_{nm}^{(1)},\lambda_{n+1m}^{(2)},\lambda_{nm+1}^{(3)},1_{n-1m+1}]}\right)
\\*[2ex]\\*[2ex] +
\displaystyle \left( \mathcal{D}_{4}- \displaystyle
\frac{\mathcal{D}_{3}\cdot
\mathcal{D}[\lambda_{nm}^{(1)},\lambda_{n+1m}^{(4)},1_{nm+1}]}{\mathcal{D}[\lambda_{nm}^{(1)},\lambda_{n+1m}^{(2)},1_{nm+1}]}
- \displaystyle
\frac{\mathcal{D}_{2}\cdot\mathcal{D}[\lambda_{nm}^{(1)},\lambda_{n+1m}^{(2)},\lambda_{nm+1}^{(4)},1_{n-1m+1}]}
{\mathcal{D}[\lambda_{nm}^{(1)},\lambda_{n+1m}^{(2)},\lambda_{nm+1}^{(3)},1_{n-1m+1}]}\right)
Z\\*[2ex]\\*[2ex] + \left(\displaystyle
\frac{\mathcal{D}_{3}\cdot\mathcal{D}[\lambda_{nm}^{(1)},\lambda_{n+1m}^{(3)},1_{nm+1}]\cdot
\mathcal{D}[\lambda_{nm}^{(1)},\lambda_{n+1m}^{(2)},\lambda_{nm+1}^{(4)},1_{n-1m+1}]}{\mathcal{D}[\lambda_{nm}^{(1)},\lambda_{n+1m}^{(2)},1_{nm+1}]\cdot
\mathcal{D}[\lambda_{nm}^{(1)},\lambda_{n+1m}^{(2)},\lambda_{nm+1}^{(3)},1_{n-1m+1}]}\right)
Z + f_{nm}(t,\xi_{pq}),\\*[2ex]\\*[2ex]
\end {array}
\label{FA69}
$$

$$
Z= \displaystyle
\frac{\mathcal{D}[u_{nm},\lambda_{n+1m}^{(1)},\lambda_{nm+1}^{(2)},\lambda_{n-1m+1}^{(3)},1_{n-1m}]}
{(\lambda_{n+1m}^{(1)}-\lambda_{nm}^{(1)})\cdot
\mathcal{D}[\lambda_{nm}^{(1)},\lambda_{n+1m}^{(2)},\lambda_{nm+1}^{(3)},\lambda_{n-1m+1}^{(4)},1_{n-1m}]},
$$

$$
\mathcal {D}_{2}=\left|%
\begin{array}{ccc}
\ddot{\lambda}_{nm}^{(1)}& \lambda_{nm}^{(1)}& \lambda_{n+1m}^{(1)}\\
\ddot{\lambda}_{nm}^{(2)}& \lambda_{nm}^{(2)}& \lambda_{n+1m}^{(2)}\\
0&1&1
\end{array}
\right|,\quad \mathcal {D}_{3}=\left|%
\begin{array}{ccc}
\ddot{\lambda}_{nm}^{(1)}& \lambda_{nm}^{(1)}& \lambda_{n+1m}^{(1)}\\
\ddot{\lambda}_{nm}^{(3)}& \lambda_{nm}^{(3)}& \lambda_{n+1m}^{(3)}\\
0&1&1
\end{array}
\right|, \quad
\mathcal {D}_{4}=\left|%
\begin{array}{ccc}
\ddot{\lambda}_{nm}^{(1)}& \lambda_{nm}^{(1)}& \lambda_{n+1m}^{(1)}\\
\ddot{\lambda}_{nm}^{(4)}& \lambda_{nm}^{(4)}& \lambda_{n+1m}^{(4)}\\
0&1&1
\end{array}
\right|,
$$
$$
\xi_{pq}=\mathcal{D}[u_{nm},\lambda_{n+1m}^{(1)},\lambda_{nm+1}^{(2)},\lambda_{n-1m+1}^{(3)},\lambda_{n-1m}^{(4)},1_{pq}],\quad
(p,q)\in \{(n,m-1),(n+1,m-1)\}, \label{InvA69}
$$
$$
\lambda_{n+1m}^{(1)} \neq \lambda_{nm}^{(1)},\quad
\mathcal{D}[\lambda_{nm}^{(1)},\lambda_{n+1m}^{(2)},1_{nm+1}]\neq
0 ,\quad
\mathcal{D}[\lambda_{nm}^{(1)},\lambda_{n+1m}^{(2)},\lambda_{nm+1}^{(3)},1_{n-1m+1}]\neq
0,
$$
$$
\mathcal{D}[\lambda_{nm}^{(1)},\lambda_{n+1m}^{(2)},
\lambda_{nm+1}^{(3)},\lambda_{n-1m+1}^{(4)},1_{n-1m}]\neq 0,\
\mbox{with }\displaystyle
\sum_{j=1}^4\Big(\ddot{\omega}_{kj}\lambda_{nm}^{(j)}+2\dot{\omega}_{kj}\dot{\lambda}_{nm}^{(j)}
\Big)+\ddot{\sigma}^{(k)}=0,\quad \det(\dot\omega_{kj})\neq 0.
$$
\end{thm}

\begin{thm}
\label{A12}
Equation (\ref{equation}) allows a $12$-dimensional abelian
symmetry algebra for $2$ classes of interactions $F_{nm}$. The
algebras and interaction functions can be represented as follows:
\begin{equation}
\label{Alg12}
A_{12,1}:\quad \widehat
X_i=\lambda_{nm}^{(i)}(t)\partial_{u_{nm}},\ i=1,\ldots,6; \quad
\widehat Y_k=\Big(\displaystyle \sum_{j=1}^6
\omega_{kj}(t)\lambda_{nm}^{(j)}(t)\Big)\partial_{u_{nm}},\ k=1,\ldots,6
\end{equation}
\begin{equation}
\begin{array}{l}
F_{nm}=1 +
\left|%
\begin{array}{ccccccc}
  u_{nm} & 1 & u_{n+1m} & u_{nm+1} & u_{n-1m+1}& u_{n-1m} &  u_{nm-1}\\
  \lambda_{nm}^{(1)} & \ddot{\lambda}_{nm}^{(1)} & \lambda_{n+1m}^{(1)} &  \lambda_{nm+1}^{(1)} &  \lambda_{n-1m+1}^{(1)} & \lambda_{n-1m}^{(1)} & \lambda_{nm-1}^{(1)}\\
  \lambda_{nm}^{(2)} & \ddot{\lambda}_{nm}^{(2)} & \lambda_{n+1m}^{(2)} &  \lambda_{nm+1}^{(2)} &  \lambda_{n-1m+1}^{(2)}& \lambda_{n-1m}^{(2)} & \lambda_{nm-1}^{(2)}\\
\lambda_{nm}^{(3)} & \ddot{\lambda}_{nm}^{(3)} & \lambda_{n+1m}^{(3)} & \lambda_{nm+1}^{(3)} & \lambda_{n-1m+1}^{(3)} &\lambda_{n-1m}^{(3)} & \lambda_{nm-1}^{(3)}\\
\lambda_{nm}^{(4)} & \ddot{\lambda}_{nm}^{(4)} & \lambda_{n+1m}^{(4)} &  \lambda_{nm+1}^{(4)} & \lambda_{n-1m+1}^{(4)} & \lambda_{n-1m}^{(4)} & \lambda_{nm-1}^{(4)}\\
\lambda_{nm}^{(5)} & \ddot{\lambda}_{nm}^{(5)} & \lambda_{n+1m}^{(5)} &  \lambda_{nm+1}^{(5)} & \lambda_{n-1m+1}^{(5)} & \lambda_{n-1m}^{(5)} & \lambda_{nm-1}^{(5)}\\
\lambda_{nm}^{(6)} & \ddot{\lambda}_{nm}^{(6)} & \lambda_{n+1m}^{(6)} &  \lambda_{nm+1}^{(6)} & \lambda_{n-1m+1}^{(6)} & \lambda_{n-1m}^{(6)} & \lambda_{nm-1}^{(6)}
\end{array}%
\right|\\*[2ex] \\*[2ex]  \times
\left(\mathcal{D}[\lambda_{nm}^{(1)},\lambda_{n+1m}^{(2)},\lambda_{nm+1}^{(3)},\lambda_{n-1m+1}^{(4)},\lambda_{n-1m}^{(5)},,\lambda_{nm-1}^{(6)}]\right)^{-1}
+\ f_{nm}(t,\xi),
\end{array}
\label{FA64}
\end{equation}
$$
\xi=\mathcal{D}[u_{nm},\lambda_{n+1m}^{(1)},\lambda_{nm+1}^{(2)},\lambda_{n-1m+1}^{(3)},\lambda_{n-1m}^{(4)},\lambda_{nm-1}^{(5)},\lambda_{n+1m-1}^{(6)}],\quad
\label{InvA64}
$$

$$
\mathcal{D}[\lambda_{nm}^{(1)},\lambda_{n+1m}^{(2)},\lambda_{nm+1}^{(3)},\lambda_{n-1m+1}^{(4)},\lambda_{n-1m}^{(5)},\lambda_{nm-1}^{(6)}]\neq 0,\ \mathcal{D}[\lambda_{nm}^{(1)},\lambda_{n+1m}^{(2)},\lambda_{nm+1}^{(3)},\lambda_{n-1m+1}^{(4)},\lambda_{n-1m}^{(5)}]\neq 0,
$$
$$
\mathcal{D}[\lambda_{nm}^{(1)},\lambda_{n+1m}^{(2)},\lambda_{nm+1}^{(3)},\lambda_{n-1m+1}^{(4)}]\neq 0,\quad
\mathcal{D}[\lambda_{nm}^{(1)},\lambda_{n+1m}^{(2)},\lambda_{nm+1}^{(3)}]\neq 0,\quad
\mathcal{D}[\lambda_{nm}^{(1)},\lambda_{n+1m}^{(2)}]\neq 0.
$$
\begin{equation}
\label{conda121}
\mbox{ with }\displaystyle
\sum_{j=1}^6\Big(\ddot{\omega}_{kj}\lambda_{nm}^{(j)}+2\dot{\omega}_{kj}\dot{\lambda}_{nm}^{(j)}
\Big)=0,\quad \det(\dot\omega_{kj})\neq 0.
\end{equation}

$$
A_{12,2}: \widehat X_1=\partial_{u_{nm}},\quad \widehat X_2=
t\partial_{u_{nm}},\quad \widehat
X_{i+2}=\lambda_{nm}^{(i)}(t)\partial_{u_{nm}},\ i=1,\ldots,5
$$
$$
\widehat Y_k=\Big(\displaystyle \sum_{j=1}^5
\omega_{kj}(t)\lambda_{nm}^{(j)}(t)+\sigma_k(t)\Big)\partial_{u_{nm}},\ k=1,\ldots,5
$$

$$
\begin{array}{l}
F_{nm}= \displaystyle
\frac{\ddot{\lambda}_{nm}^{(1)}}{\lambda_{n+1m}^{(1)}-\lambda_{nm}^{(1)}}(u_{n+1m}-u_{nm})-\displaystyle
\frac{\mathcal{D}_{3}\cdot\mathcal{D}[u_{nm},\lambda_{n+1m}^{(1)},1_{nm+1}]}{(\lambda_{n+1m}^{(1)}-
\lambda_{nm}^{(1)})\cdot\mathcal{D}[\lambda_{nm}^{(1)},\lambda_{n+1m}^{(2)},1_{nm+1}]}\\*[2ex]\\*[2ex]
+ \displaystyle \left(\mathcal{D}_{2} - \displaystyle
\frac{\mathcal{D}_{3}\cdot\mathcal{D}[\lambda_{nm}^{(1)},\lambda_{n+1m}^{(3)},1_{nm+1}]}
{\mathcal{D}[\lambda_{nm}^{(1)},\lambda_{n+1m}^{(2)},1_{nm+1}]}\right)
\displaystyle \left(
\frac{\mathcal{D}[u_{nm},\lambda_{n+1m}^{(1)},\lambda_{nm+1}^{(2)},1_{n-1m+1}]}
{(\lambda_{n+1m}^{(1)}-\lambda_{nm}^{(1)})\cdot\mathcal{D}[\lambda_{nm}^{(1)},\lambda_{n+1m}^{(2)},\lambda_{nm+1}^{(3)},1_{n-1m+1}]}\right)
\\*[2ex]\\*[2ex] + \displaystyle \left( \mathcal{D}_{4}- \displaystyle \frac{\mathcal{D}_{3}\cdot
\mathcal{D}[\lambda_{nm}^{(1)},\lambda_{n+1m}^{(4)},1_{nm+1}]}
{\mathcal{D}[\lambda_{nm}^{(1)},\lambda_{n+1m}^{(2)},1_{nm+1}]} -
\displaystyle \frac{\mathcal{D}_{2}\cdot
\mathcal{D}[\lambda_{nm}^{(1)},\lambda_{n+1m}^{(2)},\lambda_{nm+1}^{(4)},1_{n-1m+1}]}
{\mathcal{D}[\lambda_{nm}^{(1)},\lambda_{n+1m}^{(2)},\lambda_{nm+1}^{(3)},1_{n-1m+1}]}\right)
Z\\*[2ex]\\*[2ex] + \left(\displaystyle \frac{\mathcal{D}_{3}\cdot
\mathcal{D}[\lambda_{nm}^{(1)},\lambda_{n+1m}^{(3)},1_{nm+1}]\cdot
\mathcal{D}[\lambda_{nm}^{(1)},\lambda_{n+1m}^{(2)},\lambda_{nm+1}^{(4)},1_{n-1m+1}]}
{\mathcal{D}[\lambda_{nm}^{(1)},\lambda_{n+1m}^{(2)},1_{nm+1}]\cdot
\mathcal{D}[\lambda_{nm}^{(1)},\lambda_{n+1m}^{(2)},\lambda_{nm+1}^{(3)},1_{n-1m+1}]}\right)
Z \\*[2ex]\\*[2ex]
 + \displaystyle \left( \mathcal{D}_{4}- \displaystyle \frac{\mathcal{D}_{3}\cdot
 \mathcal{D}[\lambda_{nm}^{(1)},\lambda_{n+1m}^{(4)},1_{nm+1}]}
 {\mathcal{D}[\lambda_{nm}^{(1)},\lambda_{n+1m}^{(2)},1_{nm+1}]} -
 \displaystyle \frac{\mathcal{D}_{2}\cdot
 \mathcal{D}[\lambda_{nm}^{(1)},\lambda_{n+1m}^{(2)},\lambda_{nm+1}^{(4)},1_{n-1m+1}]}
 {\mathcal{D}[\lambda_{nm}^{(1)},\lambda_{n+1m}^{(2)},\lambda_{nm+1}^{(3)},1_{n-1m+1}]}\right)
 Z_{3}\\*[2ex]\\*[2ex]+ \left(\displaystyle \frac{\mathcal{D}_{3}\cdot
 \mathcal{D}[\lambda_{nm}^{(1)},\lambda_{n+1m}^{(3)},1_{nm+1}]\cdot
 \mathcal{D}[\lambda_{nm}^{(1)},\lambda_{n+1m}^{(2)},\lambda_{nm+1}^{(4)},1_{n-1m+1}]}
 {\mathcal{D}[\lambda_{nm}^{(1)},\lambda_{n+1m}^{(2)},1_{nm+1}]\cdot
 \mathcal{D}[\lambda_{nm}^{(1)},\lambda_{n+1m}^{(2)},\lambda_{nm+1}^{(3)},1_{n-1m+1}]}\right)
 Z_{3}\\*[2ex]\\*[2ex] - \left(\mathcal{D}_{2}-\displaystyle
\frac{\mathcal{D}_{3}\cdot\mathcal{D}[\lambda_{nm}^{(1)},\lambda_{n+1m}^{(3)},1_{nm+1}]}
{\mathcal{D}[\lambda_{nm}^{(1)},\lambda_{n+1m}^{(2)},1_{nm+1}]}
\right)Z_{2}+ \left( \mathcal{D}_{5}- \displaystyle
\frac{\mathcal{D}_{1}\cdot\mathcal{D}[\lambda_{nm}^{(1)},\lambda_{n+1m}^{(5)},1_{nm+1}]}
{\mathcal{D}[\lambda_{nm}^{(1)},\lambda_{n+1m}^{(2)},1_{nm+1}]}
\right)Z_{1}+ f_{nm}(t,\xi),\\*[2ex]\\*[2ex]
\end {array}
\label{FA78}
$$
$$
\xi=\mathcal{D}[u_{nm},\lambda_{n+1m}^{(1)},\lambda_{nm+1}^{(2)},\lambda_{n-1m+1}^{(3)},\lambda_{n-1m}^{(4)},\lambda_{nm-1}^{(5)},1_{n+1m-1}].
\label{InvA78}
$$
$$
Z= \displaystyle
\frac{\mathcal{D}[u_{nm},\lambda_{n+1m}^{(1)},\lambda_{nm+1}^{(2)},\lambda_{n-1m+1}^{(3)},1_{n-1m}]}{(\lambda_{n+1m}^{(1)}-\lambda_{nm}^{(1)})\cdot\mathcal{D}[\lambda_{nm}^{(1)},\lambda_{n+1m}^{(2)},\lambda_{nm+1}^{(3)},\lambda_{n-1m+1}^{(4)},1_{n-1m}]},
$$
$$
Z_{1}= \displaystyle
\frac{\mathcal{D}[u_{nm},\lambda_{n+1m}^{(1)},\lambda_{nm+1}^{(2)},\lambda_{n-1m+1}^{(3)},
\lambda_{n-1m}^{(4)},1_{nm-1}]}{(\lambda_{n+1m}^{(1)}-\lambda_{nm}^{(1)})\cdot\mathcal{D}[\lambda_{nm}^{(1)},\lambda_{n+1m}^{(2)},\lambda_{nm+1}^{(3)},\lambda_{n-1m+1}^{(4)},\lambda_{n-1m}^{(5)},1_{nm-1}]},
$$
$$
Z_{2}=
\mathcal{D}[\lambda_{nm}^{(1)},\lambda_{n+1m}^{(2)},\lambda_{nm+1}^{(5)},1_{n-1m+1}]\cdot
Z_{1}, \quad
Z_{3}=\mathcal{D}[\lambda_{nm}^{(1)},\lambda_{n+1m}^{(2)},\lambda_{nm+1}^{(3)},\lambda_{n-1m+1}^{(5)},1_{n-1m}]\cdot
Z_{1},
$$
$$
\mathcal {D}_{2}=\left|%
\begin{array}{ccc}
\ddot{\lambda}_{nm}^{(1)}& \lambda_{nm}^{(1)}& \lambda_{n+1m}^{(1)}\\
\ddot{\lambda}_{nm}^{(2)}& \lambda_{nm}^{(2)}& \lambda_{n+1m}^{(2)}\\
0&1&1
\end{array}
\right|,\quad \mathcal {D}_{3}=\left|%
\begin{array}{ccc}
\ddot{\lambda}_{nm}^{(1)}& \lambda_{nm}^{(1)}& \lambda_{n+1m}^{(1)}\\
\ddot{\lambda}_{nm}^{(3)}& \lambda_{nm}^{(3)}& \lambda_{n+1m}^{(3)}\\
0&1&1
\end{array}
\right|, \quad
\mathcal {D}_{4}=\left|%
\begin{array}{ccc}
\ddot{\lambda}_{nm}^{(1)}& \lambda_{nm}^{(1)}& \lambda_{n+1m}^{(1)}\\
\ddot{\lambda}_{nm}^{(4)}& \lambda_{nm}^{(4)}& \lambda_{n+1m}^{(4)}\\
0&1&1
\end{array}
\right|,
$$
$$
\mathcal {D}_{5}=\left|%
\begin{array}{ccc}
\ddot{\lambda}_{nm}^{(1)}& \lambda_{nm}^{(1)}& \lambda_{n+1m}^{(1)}\\
\ddot{\lambda}_{nm}^{(5)}& \lambda_{nm}^{(5)}& \lambda_{n+1m}^{(5)}\\
0&1&1
\end{array}
\right|, \quad \lambda_{n+1m}^{(1)} \neq  \lambda_{nm}^{(1)},\quad
\mathcal{D}[\lambda_{nm}^{(1)},\lambda_{n+1m}^{(2)},1_{nm+1}]\neq
0,
$$
$$
\mathcal{D}[\lambda_{nm}^{(1)},\lambda_{n+1m}^{(2)},\lambda_{nm+1}^{(3)},1_{n-1m+1}]\neq
0,\quad
\mathcal{D}[\lambda_{nm}^{(1)},\lambda_{n+1m}^{(2)},\lambda_{nm+1}^{(3)},\lambda_{n-1m+1}^{(4)},1_{n-1m}]\neq
0,
$$
$$
\mathcal{D}[\lambda_{nm}^{(1)},\lambda_{n+1m}^{(2)},\lambda_{nm+1}^{(3)},\lambda_{n-1m+1}^{(4)},
\lambda^{(5)}_{n-1m},1_{nm-1}]\neq 0,\  \mbox{with }\displaystyle
\sum_{j=1}^5\Big(\ddot{\omega}_{kj}\lambda_{nm}^{(j)}+2\dot{\omega}_{kj}\dot{\lambda}_{nm}^{(j)}
\Big)+\ddot{\sigma}^{(k)}=0,\ \det(\dot\omega_{kj})\neq 0.
$$

%
%
%$$
%A_{8,10}: \widehat X_1=\partial_{u_{nm}},\quad \widehat X_2=t
%\partial_{u_{nm}}, \quad \widehat
%X_3=\lambda_{nm}(t)\partial_{u_{nm}},\quad \widehat
%X_4=\Big(\omega_1(t)\lambda_{nm}(t)+\omega_2(t)\Big)\partial_{u_{nm}}
%$$
%$$
%\widehat
%X_5=\Big(\omega_3(t)\lambda_{nm}(t)+\omega_4(t)\Big)\partial_{u_{nm}},\quad
%\widehat
%X_6=\Big(\omega_5(t)\lambda_{nm}(t)+\omega_6(t)\Big)\partial_{u_{nm}},
%$$

%$$
%\widehat
%X_7=\Big(\omega_7(t)\lambda_{nm}(t)+\omega_8(t)\Big)\partial_{u_{nm}},\quad
%\widehat
%X_8=\Big(\omega_9(t)\lambda_{nm}(t)+\omega_{10}(t)\Big)\partial_{u_{nm}},
%$$
%$$
%\mbox{with }\ \
%\ddot{\omega}_{k+1}\lambda_{nm}+2\dot{\omega}_{k+1}\dot{\lambda}_{nm}+\ddot{\omega}_{k+2}=0\mbox{
%for }k=0,2,4,6,8
%$$
%$$
%\quad F_{nm} \mbox{ as in (\ref{FA36}) and (\ref{InvA36}),with the
%same conditions.} $$
\end{thm}

\section{Nonsolvable symmetry algebras}
The vector fields of equation (\ref{equation}) should have the form (\ref{vector2}). We can now ask whether it is possible to obtain simple symmetry algebras from these vector fields. We obtain the following theorem.

\begin{thm}
Equation $(\ref{equation})$ allows only one simple Lie algebra, $\mathrm{sl}(2,\mathbb{R})$, given by:
\begin{equation}
\label{sl2r}
NS_{3,1}: \widehat X_1=\partial_{t}, \quad \widehat X_2=t\partial_{t}+ \frac{1}{2}u_{nm}\partial_{u_{nm}}, \quad \widehat X_3=t^{2}\partial_{t}+ tu_{nm}\partial_{u_{nm}},
\end{equation}
$$
F_{nm}= \frac{1}{u_{nm}^{3}}f_{nm}(\xi_{pq}), \quad \xi_{pq}=\frac{u_{pq}}{u_{nm}}, \quad (p,q)\in\Gamma \setminus\{(n,m)\},
$$
$$
[\widehat X_1,\widehat X_2]=\widehat X_1,\ \ \ \ [\widehat X_1,\widehat X_3]=2\widehat X_2,\ \ \ \ [\widehat X_2,\widehat X_3]=\widehat X_3.
$$
\end{thm}
We can now look for additional symmetries by considering nonsolvable symmetry algebras for equation $(\ref{equation})$. A nonsovable Lie algebra must contain a simple subalgebra, i.e. the Lie algebra  $\mathrm{sl}(2,\mathbb{R})$ of $NS_{3,1}$ in our case. Therefore, we add new vector fields $\widehat Y_i$ of the form (\ref{vector2}) to $NS_{3,1}$. These vector fields $\{\widehat Y_i\}$ forming the radical of the new Lie algebras. The following theorems give the possible nonsovable symmetry Lie algebras for equation $(\ref{equation})$. We only list the radical of the nonsolvable Lie algebras since all of these algebras have $\mathrm{sl}(2,\mathbb{R})$ of the form (\ref{sl2r}) as subalgebra.
\begin{thm}
Equation $(\ref{equation})$ allows a $4$-dimensional nonsolvable symmetry algebra for $1$ classe of interaction:
$$
NS_{4,1}: \widehat Y_1=a_{nm}u_{nm}\partial_{u_{nm}},\quad F_{nm}=u_{nm}\left[\big(u_{nm}\big)^{a_{n+1m}}\big(u_{n+1m}\big)^{-a_{nm}}\right]^{\frac{4}{\mathcal{D}[a_{nm},1_{n+1m}]}}f_{nm}(\xi_{pq}),
$$
$$
\xi_{pq}=(u_{nm})^{-\mathcal{D}[a_{n+1m},1_{pq}]}(u_{n+1m})^{\mathcal{D}[a_{nm},1_{pq}]}(u_{pq})^{-\mathcal{D}[a_{nm},1_{n+1m}]},\quad (p,q)\in\Gamma \setminus\{(n,m),(n+1,m)\}.
$$
\end{thm}

\begin{thm}
Equation $(\ref{equation})$ allows a $5$-dimensional nonsolvable symmetry algebra for $2$ classes of interactions:
$$
NS_{5,1} : \widehat Y_1=a_{nm}^{(1)}u_{nm}\partial_{u_{nm}}, \quad \widehat Y_2=a_{nm}^{(2)}u_{nm}\partial_{u_{nm}}
$$
$$
\begin{array}{rcl}
F_{nm}&=&u_{nm}\left[\big(u_{nm}\big)^{-\mathcal{D}[a_{n+1m}^{(1)},a_{nm+1}^{(2)}]}\big(u_{n+1m}\big)^{\mathcal{D}[a_{nm}^{(1)},a_{nm+1}^{(2)}]}
\big(u_{nm+1}\big)^{-\mathcal{D}[a_{nm}^{(1)},a_{n+1m}^{(2)}]}\right]^{\frac{4}{\mathcal{D}[a_{nm}^{(1)},a_{n+1m}^{(2)},1_{nm+1}]}}\\*[2ex] && \times f_{nm}(\xi_{pq}),\quad (p,q)\in\Gamma \setminus\{(n,m),(n+1,m),(nm+1)\},
\end{array}
$$
$$
\xi_{pq}=(u_{nm})^{\mathcal{D}[a_{n+1m}^{(1)},a_{nm+1}^{(2)},1_{pq}]}(u_{n+1m})^{-\mathcal{D}[a_{nm}^{(1)},a_{nm+1}^{(2)},1_{pq}]}(u_{nm+1})^{\mathcal{D}[a_{nm}^{(1)},a_{n+1m}^{(2)},1_{pq}]}(u_{pq})^{-\mathcal{D}[a_{nm}^{(1)},a_{n+1m}^{(2)},1_{nm+1}]}.
$$

$$
NS_{5,2}: \widehat Y_1=\partial_{u_{nm}},\quad \widehat Y_2=t\partial_{u_{nm}}, \quad F_{nm}= \big(\mathcal{D}[u_{nm},1_{n+1m}]\big)^{-3}f_{nm}(\xi_{pq})
$$
$$
\xi_{pq}=\frac{\mathcal{D}[u_{nm},1_{pq}]}{\mathcal{D}[u_{nm},1_{n+1m}]},\quad (p,q)\in\Gamma \setminus\{(n,m),(n+1,m)\}.
$$
\end{thm}

\begin{thm}
Equation $(\ref{equation})$ allows a $6$-dimensional nonsolvable symmetry algebra for $1$ classe of interaction:
$$
NS_{6,1}: \widehat Y_1=a_{nm}^{(1)}u_{nm}\partial_{u_{nm}},\quad \widehat Y_2=a_{nm}^{(2)}u_{nm}\partial_{u_{nm}}, \quad \widehat Y_3=a_{nm}^{(3)}u_{nm}\partial_{u_{nm}}
$$
$$
\begin{array}{rcl}
F_{nm}&=&u_{nm}\left[\big(u_{nm}\big)^{\mathcal{D}[a_{n+1m}^{(1)},a_{nm+1}^{(2)},a_{n-1m+1}^{(3)}]}
\big(u_{n+1m}\big)^{-\mathcal{D}[a_{nm}^{(1)},a_{nm+1}^{(2)},a_{n-1m+1}^{(3)}]}
\big(u_{nm+1}\big)^{\mathcal{D}[a_{nm}^{(1)},a_{n+1m}^{(2)},a_{n-1m+1}^{(3)}]} \right. \\[2ex] && \times \left. \big(u_{n-1m+1}\big)^{-\mathcal{D}[a_{nm}^{(1)},a_{n+1m}^{(2)},a_{nm+1}^{(3)}]}\right]^{\alpha_{nm}} f_{nm}(\xi_{pq}),
\end{array}
$$
$$
\alpha_{nm}={\frac{4}{\mathcal{D}[a_{nm}^{(1)},a_{n+1m}^{(2)},a_{nm+1}^{(3)},1_{n-1m+1}]}}, \quad (p,q) \in \{(n-1,m),(n,m-1),(n+1,m-1)\}.
$$
$$
\begin{array}{rcl}
\xi_{pq}&=&\big(u_{nm}\big)^{-\mathcal{D}[a_{n+1m}^{(1)},a_{nm+1}^{(2)},a_{n-1m+1}^{(3)},1_{pq}]}
\big(u_{n+1m}\big)^{\mathcal{D}[a_{nm}^{(1)},a_{nm+1}^{(2)},a_{n-1m+1}^{(3)},1_{pq}]}\\[2ex] && \times
\big(u_{nm+1}\big)^{-\mathcal{D}[a_{nm}^{(1)},a_{n+1m}^{(2)},a_{n-1m+1}^{(3)},1_{pq}]}
\big(u_{n-1m+1}\big)^{\mathcal{D}[a_{nm}^{(1)},a_{n+1m}^{(2)},a_{nm+1}^{(3)},1_{pq}]}\\[2ex] && \times \big(u_{pq}\big)^{-\mathcal{D}[a_{nm}^{(1)},a_{n+1m}^{(2)},a_{nm+1}^{(3)},1_{n-1m+1}]}.
\end{array}
$$
\end{thm}

\begin{thm}
Equation $(\ref{equation})$ allows a $7$-dimensional nonsolvable symmetry algebra for $2$ classes of interactions:

$$
NS_{7,1}: Y_i=a_{nm}^{(i)}u_{nm}\partial_{u_{nm}}, \quad i=1,\ldots ,4; \quad (p,q) \in \{(n,m-1),(n+1,m-1)\},
$$

$$
\begin{array}{rcl}
F_{nm}&=& u_{nm}\left[(u_{nm})^{-\mathcal{D}[a_{n+1m}^{(1)},a_{nm+1}^{(2)},a_{n-1m+1}^{(3)},a_{n-1m}^{(4)}]}(u_{n+1m})^{\mathcal{D}[a_{nm}^{(1)},a_{nm+1}^{(2)},a_{n-1m+1}^{(3)},a_{n-1m}^{(4)}]}\right.\\[2ex]
&& \times (u_{nm+1})^{-\mathcal{D}[a_{nm}^{(1)},a_{n+1m}^{(2)},a_{n-1m+1}^{(3)},a_{n-1m}^{(4)}]}(u_{n-1m+1})^{\mathcal{D}[a_{nm}^{(1)},a_{n+1m}^{(2)},a_{nm+1}^{(3)},a_{n-1m}^{(4)}]}\\[2ex]
&& \times \left.(u_{n-1m})^{-\mathcal{D}[a_{nm}^{(1)},a_{n+1m}^{(2)},a_{nm+1}^{(3)},a_{n-1m+1}^{(4)}]}\right]^{\alpha_{nm}} f_{nm}(\xi_{pq}),
\end{array}
$$

$$
\begin{array}{rcl}
\xi_{pq}&=& (u_{nm})^{\mathcal{D}[a_{n+1m}^{(1)},a_{nm+1}^{(2)},a_{n-1m+1}^{(3)},a_{n-1m}^{(4)},1_{pq}]}(u_{n+1m})^{-\mathcal{D}[a_{nm}^{(1)},a_{nm+1}^{(2)},a_{n-1m+1}^{(3)},a_{n-1m}^{(4)},1_{pq}]}\\[2ex]&& \times (u_{nm+1})^{\mathcal{D}[a_{nm}^{(1)},a_{n+1m}^{(2)},a_{n-1m+1}^{(3)},a_{n-1m}^{(4)},1_{pq}]}(u_{n-1m+1})^{-\mathcal{D}[a_{nm}^{(1)},a_{n+1m}^{(2)},a_{nm+1}^{(3)},a_{n-1m}^{(4)},1_{pq}]} \\[2ex]&& \times (u_{n-1m})^{\mathcal{D}[a_{nm}^{(1)},a_{n+1m}^{(2)},a_{nm+1}^{(3)},a_{n-1m+1}^{(4)},1_{pq}]}(u_{pq})^{-\mathcal{D}[a_{nm}^{(1)},a_{n+1m}^{(2)},a_{nm+1}^{(3)},a_{n-1m+1}^{(4)},1_{n-1m}]},
\end{array}
$$

$$
\alpha_{nm}= \frac{4}{\mathcal{D}[a_{nm}^{(1)},a_{n+1m}^{(2)},a_{nm+1}^{(3)},a_{n-1m+1}^{(4)},1_{n-1m}]}.
$$

$$
NS_{7,2}: \widehat Y_1=\partial_{u_{nm}},\quad \widehat Y_2= t\partial_{u_{nm}}, \quad \widehat Y_3= \kappa_{nm} \partial_{u_{nm}}, \quad \widehat Y_4= \kappa_{nm}t\partial_{u_{nm}},
$$
$$
 \quad \dot{\kappa}_{nm}=0, \quad \kappa_{nm}\neq \kappa_{n+1m}, \quad \kappa_{nm}\neq \kappa_{nm+1},
$$
$$
F_{nm}=\big(\mathcal{D}[u_{nm},\kappa_{n+1m},1_{nm+1}]\big)^{-3}f_{nm}(\xi_{pq}), \quad \xi_{pq}= \frac{\mathcal{D}[u_{nm},\kappa_{n+1m},1_{pq}]}{\mathcal{D}[u_{nm}, \kappa_{n+1m},1_{nm+1}]},
$$
$$
\mathcal{D}[u_{nm}, \kappa_{n+1m},1_{nm+1}], \quad (p,q) \in\Gamma\setminus\{(n,m),(n+1,m),(n,m+1)\}.
$$
\end{thm}

\begin{thm}
Equation $(\ref{equation})$ allows a $8$-dimensional nonsolvable symmetry algebra for $1$ classe of interaction:
$$
NS_{8,1}: \widehat Y_i=a_{nm}^{(i)}u_{nm}\partial_{u_{nm}}, \quad i=1,\ldots,5;
$$

$$
\begin{array}{rcl}
F_{nm}&=& u_{nm}\left[(u_{nm})^{\mathcal{D}[a_{n+1m}^{(1)},a_{nm+1}^{(2)},a_{n-1m+1}^{(3)},a_{n-1m}^{(4)},a_{nm-1}^{(5)}]}(u_{n+1m})^{-\mathcal{D}[a_{nm}^{(1)},a_{nm+1}^{(2)},a_{n-1m+1}^{(3)},a_{n-1m}^{(4)},a_{nm-1}^{(5)}]}\right. \\[2ex]
&& \times (u_{nm+1})^{\mathcal{D}[a_{nm}^{(1)},a_{n+1m}^{(2)},a_{n-1m+1}^{(3)},a_{n-1m}^{(4)},a_{nm-1}^{(5)}]}(u_{n-1m+1})^{-\mathcal{D}[a_{nm}^{(1)},a_{n+1m}^{(2)},a_{nm+1}^{(3)},a_{n-1m}^{(4)},a_{nm-1}^{(5)}]}\\[2ex]
&& \times \left. (u_{n-1m})^{\mathcal{D}[a_{nm}^{(1)},a_{n+1m}^{(2)},a_{nm+1}^{(3)},a_{n-1m+1}^{(4)},a_{nm-1}^{(5)}]}(u_{nm-1})^{\mathcal{D}[a_{nm}^{(1)},a_{n+1m}^{(2)},a_{nm+1}^{(3)},a_{n-1m+1}^{(4)},a_{n-1m}^{(5)}]}\right]^{\alpha_{nm}}\\[2ex]&& \times f_{nm}(\xi),\quad \alpha_{nm}=\frac{4}{\mathcal{D}[a_{nm}^{(1)},a_{n+1m}^{(2)},a_{nm+1}^{(3)},a_{n-1m+1}^{(4)},a_{n-1m}^{(5)},1_{nm-1}]},
\end{array}
$$

$$
\begin{array}{l}
\xi= (u_{nm})^{-\mathcal{D}[a_{n+1m}^{(1)},a_{nm+1}^{(2)},a_{n-1m+1}^{(3)},a_{n-1m}^{(4)},a_{nm-1}^{(5)},1_{n+1m-1}]}(u_{n+1m})^{\mathcal{D}[a_{nm}^{(1)},a_{nm+1}^{(2)},a_{n-1m+1}^{(3)},a_{n-1m}^{(4)},a_{nm-1}^{(5)},1_{n+1m-1}]}\\[2ex]
\times (u_{nm+1})^{-\mathcal{D}[a_{nm}^{(1)},a_{n+1m}^{(2)},a_{n-1m+1}^{(3)},a_{n-1m}^{(4)},a_{nm-1}^{(5)},1_{n+1m-1}]}(u_{n-1m+1})^{\mathcal{D}[a_{nm}^{(1)},a_{n+1m}^{(2)},a_{nm+1}^{(3)},a_{n-1m}^{(4)},a_{nm-1}^{(5)},1_{n+1m-1}]} \\[2ex] \times (u_{n-1m})^{-\mathcal{D}[a_{nm}^{(1)},a_{n+1m}^{(2)},a_{nm+1}^{(3)},a_{n-1m+1}^{(4)},a_{nm-1}^{(5)},1_{n+1m-1}]}(u_{nm-1})^{\mathcal{D}[a_{nm}^{(1)},a_{n+1m}^{(2)},a_{nm+1}^{(3)},a_{n-1m+1}^{(4)},a_{n-1m}^{(5)},1_{n+1m-1}]}\\[2ex]\times(u_{n+1m-1})^{-\mathcal{D}[a_{nm}^{(1)},a_{n+1m}^{(2)},a_{nm+1}^{(3)},a_{n-1m+1}^{(4)},a_{n-1m}^{(5)},1_{nm-1}]}.
\end{array}
$$

\end{thm}

$$$$

\begin{thm}
Equation $(\ref{equation})$ allows a $9$-dimensional nonsolvable symmetry algebra for $1$ classe of interaction:
$$
\begin{array}{ll}
NS_{9,1}: & \widehat Y_1=\partial_{u_{nm}}, \quad \widehat Y_2= t\partial_{u_{nm}}, \quad \widehat Y_3= \kappa_{nm}^{(1)} \partial_{u_{nm}}, \quad \widehat Y_4= \kappa_{nm}^{(1)}t\partial_{u_{nm}}, \\*[2ex] & \widehat Y_5= \kappa_{nm}^{(2)} \partial_{u_{nm}}, \quad
\widehat Y_6= \kappa_{nm}^{(2)}t\partial_{u_{nm}},
\end{array}
$$

$$
\quad \dot{\kappa}_{nm}^{(i)}=0, \quad \kappa_{nm}^{(i)}\neq \kappa_{n+1m}^{(i)}, \quad \kappa_{nm}^{(i)}\neq \kappa_{nm+1}^{(i)},\quad i=1,2;
$$
$$
F_{nm}=\left(\mathcal{D}[u_{nm},\kappa_{n+1m}^{(1)},\kappa_{nm+1}^{(2)},1_{n-1m+1}]\right)^{-3}f_{nm}(\xi_{pq}), \quad \xi_{pq}= \frac{\mathcal{D}[u_{nm},\kappa_{n+1m}^{(1)},\kappa_{nm+1}^{(2)},1_{pq}]}{\mathcal{D}[u_{nm}, \kappa_{n+1m}^{(1)},\kappa_{nm+1}^{(2)},1_{n-1m+1}]},
$$
$$
(p,q) \in \{(n-1,m),(n,m-1),(n+1,m-1)\}.
$$

\end{thm}

\begin{thm}
\label{Nonsolvable11}
Equation $(\ref{equation})$ allows a $11$-dimensional nonsolvable symmetry algebra for $1$ classe of interaction:
\begin{equation}
\begin{array}{ll}
NS_{11,1}: &  \widehat Y_1=\partial_{u_{nm}},\quad \widehat Y_2= t\partial_{u_{nm}}, \quad \widehat Y_3= \kappa_{nm}^{(1)} \partial_{u_{nm}}, \quad \widehat Y_4= \kappa_{nm}^{(1)}t\partial_{u_{nm}}, \quad \widehat Y_5= \kappa_{nm}^{(2)} \partial_{u_{nm}}, \\*[2ex] & \widehat Y_6= \kappa_{nm}^{(2)}t\partial_{u_{nm}}, \quad \widehat Y_7= \kappa_{nm}^{(3)} \partial_{u_{nm}}, \quad \widehat Y_8= \kappa_{nm}^{(3)}t\partial_{u_{nm}},
\end{array}
\label{vector11}
\end{equation}
\begin{equation}
\dot{\kappa}_{nm}^{(i)}=0,\quad i={1,2,3};\quad F_{nm}=\left(
\mathcal{D}[u_{nm},\kappa_{n+1m}^{(1)},\kappa_{nm+1}^{(2)},\kappa_{n-1m+1}^{(3)},1_{n-1m}]\right)^{-3}f_{nm}(\xi_{pq}),
\label{functionF11}
\end{equation}
$$
\xi_{pq}= \frac{\mathcal{D}[u_{nm},\kappa_{n+1m}^{(1)},\kappa_{nm+1}^{(2)},\kappa_{n-1m+1}^{(3)},1_{pq}]}{\mathcal{D}[u_{nm}, \kappa_{n+1m}^{(1)},\kappa_{nm+1}^{(2)},\kappa_{n-1m+1}^{(3)},1_{n-1m}]},\quad (p,q) \in \{(n,m-1),(n+1,m-1)\}.
\label{variable11}
$$
\end{thm}

\begin{thm}
\label{NonSolvable13}
Equation $(\ref{equation})$ allows a $13$-dimensional nonsolvable symmetry algebra for $1$ classe of interaction:
\begin{equation}
\begin{array}{ll}
NS_{13,1}: & \widehat Y_1=\partial_{u_{nm}},\quad \widehat Y_2= t\partial_{u_{nm}}, \quad \widehat Y_3= \kappa_{nm}^{(1)} \partial_{u_{nm}}, \quad \widehat Y_4= \kappa_{nm}^{(1)}t\partial_{u_{nm}}, \quad \widehat Y_5= \kappa_{nm}^{(2)} \partial_{u_{nm}}, \\*[2ex] & \widehat Y_6= \kappa_{nm}^{(2)}t\partial_{u_{nm}},\quad \widehat Y_7= \kappa_{nm}^{(3)} \partial_{u_{nm}}, \quad \widehat Y_8= \kappa_{nm}^{(3)}t\partial_{u_{nm}},\quad \widehat Y_9= \kappa_{nm}^{(4)} \partial_{u_{nm}}, \quad \widehat Y_{10}= \kappa_{nm}^{(4)}t\partial_{u_{nm}},
\end{array}
\label{vector13}
\end{equation}
\begin{equation}
F_{nm}=\left(
\mathcal{D}[u_{nm},\kappa_{n+1m}^{(1)},\kappa_{nm+1}^{(2)},\kappa_{n-1m+1}^{(3)},\kappa_{n-1m}^{(4)},1_{nm-1}]\right)^{-3}f_{nm}(\xi),
\label{functionF13}
\end{equation}
$$
\dot{\kappa}_{nm}^{(i)}=0,\quad i=1,\ldots,4;\quad \xi= \frac{\mathcal{D}[u_{nm},\kappa_{n+1m}^{(1)},\kappa_{nm+1}^{(2)},\kappa_{n-1m+1}^{(3)},\kappa_{n-1m}^{(4)},1_{n+1m-1}]}{\mathcal{D}[u_{nm}, \kappa_{n+1m}^{(1)},\kappa_{nm+1}^{(2)},\kappa_{n-1m+1}^{(3)},\kappa_{n-1m}^{(4)},1_{nm-1}]}.
$$
\end{thm}

\section{Conclusions}
Group theoretical methods have been used to classify equation (\ref{equation}) according to their symmetry groups. Abelian and nonsolvable symmetry algebras $\mathcal{L}$ have been considered; the class of linear equations (\ref{equation}) has been excluded. The results of the symmetry classification are summed up in the following table:

\begin{center}

\begin{tabular}{|c|c|c|c|}
  \hline
  $\dim \mathcal{L}$ & Abelian & Nonsolvable & Total \\
  \hline \hline
  1 & 3 & 0 & 3 \\
  2 & 4 & 0 & 4 \\
  3 & 3 & 1 & 4 \\
  4 & 5 & 1 & 6 \\
  5 & 3 & 2 & 5 \\
  6 & 5 & 1 & 6 \\
  7 & 2 & 2 & 4 \\
  8 & 2 & 1 & 3 \\
  9 & 0 & 1 & 1 \\
  10 & 2 & 0 & 2 \\
  11 & 0 & 1 & 1 \\
  12 & 2 & 0 & 2 \\
  13 & 0 & 1 & 1 \\
  \hline
\end{tabular}
\vspace{2mm}

\textbf{Table 1.} Results of the symmetry classification of
equation (\ref{equation}).
\end{center}

Classification given in this paper has applications in solid-state
physics. The interest in two-dimensional systems has its origin in
possible applications to magnetic systems as well as to absorbed
layers. The theoretical models treated so far are mainly
spin-models and range from Ising systems with competing
interactions to planar Heisenberg-models \cite{20,21}. Models
involving equation of the form (\ref{equation}) appear in
references \cite{1,2,18,19} where analytic and numeric
calculations were performed. Lie point symmetries obtained in this
work could be considered to obtain analytical solutions of these
models by using symmetries to generate new solutions from a known
one or by using the `symmetry reduction method'. Moreover, some
interactions found in this work could be considered as models with
appropriate symmetries.

A continuation of this study is in progress. We know that the
existence of many symmetries is an indication of integrability.
Consequently we can ask ourselves which of the equations that are
completely specified by their Lie algebras, and therefore that
have many symmetries, are integrable. Completely specified
equations to be considered for abelian and nonsolvable Lie
algebras are, respectively, the following:
$$
A_{6,3},\quad A_{7,1},\quad A_{7,2},\quad A_{12,1}\quad
\mbox{and}\quad  A_{12,2},
$$
and
$$
NS_{8,1},\quad \mbox{and}\quad NS_{13,1}.
$$

Finally, a further task is to complete the classification, that is
to treat the nilpotent and the solvable Lie algebras.

\section*{Appendix}
This appendix includes the details of the proof for one abelian algebra and one nonsolvable algebra in the higher dimensional cases.

For the abelian Lie algebra we are considering the proof of $A_{12,1}$ of theorem~\ref{A12}. Since the procedure to obtain this classification is to proceed by dimension (for each type of algebras: abelian or nonsolvable), we suppose that we already have obtained the algebra $A_{10,1}$ of theorem~\ref{A10}. We add a new vector field of the form (\ref{vector2}), i.e. $\widehat
Z=\tau(t)\partial_t+\left[\left(\frac{\dot{\tau}}{2}+a_{nm}\right)u_{nm}+\lambda_{nm}(t)\right]\partial_{u_{nm}}$, to the symmetry algebra (\ref{alg10}).

Considering the commutation relations $[\widehat X_i,\widehat Z]=0$ and $[\widehat Y_i,\widehat Z]=0$ for $i=1,\ldots,5$ we obtain
\begin{equation}
\label{implcommabel}
\tau\dot{\lambda}^{(i)}_{nm}=\lambda_{nm}^{(i)}\Big(\frac{\dot{\tau}}{2}+a_{nm}\Big) \quad\quad \mbox{ and }\quad\quad \tau\Big(\displaystyle \sum_{j=1}^5
\dot{\omega}_{ij}\lambda_{nm}^{(j)}+\omega_{kj}\dot{\lambda}_{nm}^{(j)}\Big)=\Big(\displaystyle \sum_{j=1}^5
\omega_{ij}\lambda_{nm}^{(j)}\Big)\Big(\frac{\dot{\tau}}{2}+a_{nm}\Big).
\end{equation}
We separate the proof in two cases:
\begin{enumerate}
\item[A)] The case $\tau=0$. From the first preceding equations we easily find $\widehat
Z=\lambda^{(6)}_{nm}(t)\partial_{u_{nm}}$, where $\lambda_{nm}:=\lambda^{(6)}_{nm}(t)$ for convenience. We want now to solve the remaining determining equation (\ref{remaining}):
$$
\ddot{\lambda}_{nm}^{(6)}=\displaystyle \sum_{(p,q)\in \Gamma} \lambda_{pq}^{(6)}\partial_{u_{pq}}F_{nm}
$$
where $F_{nm}$ is the interaction (\ref{FA54}) of $A_{10,1}$. This equation is equivalent to
\begin{equation}
\label{equirem}
\begin{array}{ll}
& \mathcal{D}[\lambda^{(1)}_{nm},\lambda^{(2)}_{n+1m},\lambda^{(3)}_{nm+1},\lambda^{(4)}_{n-1m+1},\lambda^{(5)}_{n-1m}] \\*[2ex]
\times & \displaystyle \sum_{(p,q)\in \Gamma'} \mathcal{D}[\lambda^{(1)}_{nm},\lambda^{(2)}_{n+1m},\lambda^{(3)}_{nm+1},\lambda^{(4)}_{n-1m+1},\lambda^{(5)}_{n-1m},\lambda^{(6)}_{pq}]
\partial_{\xi_{pq}}f(t,\xi_{pq}) \\*[2ex]
+& \mathcal{D}[\ddot{\lambda}^{(1)}_{nm},\lambda^{(2)}_{nm},\lambda^{(3)}_{n+1m},\lambda^{(4)}_{nm+1},\lambda^{(5)}_{n-1m+1},\lambda^{(6)}_{n-1m}]=0,
\end{array}
\end{equation}
where $\Gamma':=\{(n,m-1),\ (n+1,m-1)\}$. Using then the method of characteristics we find interaction (\ref{FA64}) with symmetry algebra of dimension $11$: where the vector field $\widehat Z:=\widehat X_6=\lambda^{(6)}_{nm}\partial_{u_{nm}}$ is added to the Lie algebra of $A_{10,1}$.

%\item[A.2)] Suppose $\mathcal{D}[\lambda^{(1)}_{nm},\lambda^{(2)}_{n+1m},\lambda^{(3)}_{nm+1},\lambda^{(4)}_{n-1m+1},\lambda^{(5)}_{n-1m},\lambda^{(6)}_{nm-1}]=0$. In this case we have $\lambda_{nm}^{(6)}=\sum_{j=1}^5 \omega_{6j}(t)\lambda_{nm}^{(j)}(t)$ and equation (\ref{equirem}) becomes condition (\ref {cond101}) of $A_{10,1}$. Hence, in this case the vector field $\widehat Z:=\widehat Y_6=\sum_{j=1}^5 \big(\omega_{6j}\lambda_{nm}^{(j)}\big)\partial_{u_{nm}}$ is added to the case $A_{10,1}$ with the same conditions and interaction.

\item[B)] The case $\tau\neq 0$. By the allowed transformations we choose $\tilde t$ such that $\tau \tilde t=1$ which implies $\tau=1$. Moreover, we choose $Q_{nm}(t)$ for which $a_{nm}Q_{nm}(t)+\lambda_{nm}(t)-\dot{Q}_{nm}=0$ and we obtain $\widehat Z=\partial_t+a_{nm}u_{nm}\partial_{u_{nm}}$. First equation of (\ref{implcommabel}) implies that $\lambda_{nm}^{(i)}=\kappa_{nm}^{(i)}\mathrm{e}^{a_{nm}t}$ $i=1,\ldots,5$ where $\kappa_{nm}^{(i)}$ is an arbitrary function of $n,m$. Again here, using $P_{nm}$ in the allowed transformations, we can normalize $\kappa_{nm}^{(1)}$ to $1$. Therefore, the Lie algebra we are considering contains the subalgebra $A_{6,2}$ and the classification has been already done in this lower dimension.
\end{enumerate}
We are now looking if, for the interaction (\ref{FA64}), an additional vector field $\widehat
Z=\tau(t)\partial_t+[(\frac{\dot{\tau}}{2}+a_{nm})u_{nm}+\lambda_{nm}(t)]\partial_{u_{nm}}$ can be added to the symmetry algebras obtained in the case A. The calculations are similar to those presented above. The commutation relations considered are then $[\widehat X_i,\widehat Z]=0$ and $[\widehat Y_j,\widehat Z]=0$ for $i=1,\ldots,6$ and $j=1,\ldots,5$. Remaining equation (\ref{remaining}) implies:
\begin{equation}
\label{equirem}
\begin{array}{ll}
& \mathcal{D}[\lambda^{(1)}_{nm},\lambda^{(2)}_{n+1m},\lambda^{(3)}_{nm+1},\lambda^{(4)}_{n-1m+1},\lambda^{(5)}_{n-1m},\lambda^{(6)}_{nm-1}]\\*[2ex]
\times & \displaystyle \sum_{(p,q)\in \Gamma'} \mathcal{D}[\lambda^{(1)}_{nm},\lambda^{(2)}_{n+1m},\lambda^{(3)}_{nm+1},\lambda^{(4)}_{n-1m+1},\lambda^{(5)}_{n-1m},\lambda^{(6)}_{nm-1},
\lambda^{(7)}_{n+1m-1}]
\partial_{\xi}f(t,\xi) \\*[2ex]
+ & \mathcal{D}[\ddot{\lambda}^{(1)}_{nm},\lambda^{(2)}_{nm},\lambda^{(3)}_{n+1m},\lambda^{(4)}_{nm+1},\lambda^{(5)}_{n-1m+1},\lambda^{(6)}_{n-1m},\lambda^{(7)}_{nm-1}]=0,
\end{array}
\end{equation}
where $\lambda_{nm}:=\lambda_{nm}^{(7)}(t)$ and $F_{nm}$ is given by interaction (\ref{FA64}). Interaction (\ref{FA64}) is invariant when $\widehat Z$ is added to algebra of A if $\lambda_{nm}^{(7)}(t)=\sum_{j=1}^6 \omega_{6j}(t)\lambda_{nm}^{(j)}(t)$. Equation (\ref{equirem}) then becomes equivalent to condition (\ref{conda121}) of $A_{12,1}$. Hence, in this case the vector field $\widehat Z:=\widehat Y_6=\big(\sum_{j=1}^6 \omega_{6j}\lambda_{nm}^{(j)}\big)\partial_{u_{nm}}$ is added to the vector field of the case A with interaction (\ref{FA64}). $\Box$

We are now considering the details of the proof for the highest dimensional nonsolvable case, i.e. for theorem~\ref{NonSolvable13}. We suppose that we already know classification for dimension $11$, i.e.  theorem~\ref{Nonsolvable11} has already been shown. We add a new vector field of the form $\widehat Y_9=\tau(t)\partial_t+\left[\left(\frac{\dot{\tau}}{2}+a_{nm}\right)u_{nm}+\lambda_{nm}(t)\right]\partial_{u_{nm}}$ to the vector fields (\ref{vector11}). The Levi theorem \cite{22,23} tells us that every finite-dimensional
Lie algebra $\mathcal{L}$ is a semidirect sum of a semisimple Lie algebra $S$ and a solvable ideal (the
radical $R$):
$$
\mathcal{L}=S\rhd R,\quad \quad \quad[S,S]=S,\quad\quad \quad [S,R]\subseteq R,\quad\quad\quad [R,R]\subset R,
$$
such that we have
$$
[\widehat X_i,\widehat Y_9]=\sum_{k=1}^9 \alpha_{ik} \widehat Y_k,\quad i=1,2,3 \quad\quad \mbox{ and } \quad\quad [\widehat Y_j,\widehat Y_9]=\sum_{k=1}^9 \beta_{jk} \widehat Y_k, \quad j=1,\ldots,8
$$
where $\alpha_{ik},\beta_{jk}$ are real constants. The commutation relation $[\widehat X_1,\widehat Y_9]$ gives us that

$$
\tau(t)=\tau_0\mathrm{e}^{\alpha_{19}t},\quad \alpha_{19}a_{nm}=0,\quad \lambda_{nm}(t)=
\left\{
\begin{array}{ll}
\kappa^{(4)}_{nm}\mathrm{e}^{\alpha_{19}t}-\frac{1}{\alpha_{19}^2}\big(\alpha_{19}A^{(1)}_{nm}+\alpha_{19}B^{(1)}_{nm}t+B^{(1)}_{nm}\big),& \alpha_{19}\neq 0, \\*[2ex]
\frac{1}{2}B^{(1)}_{nm}t^2+A^{(1)}_{nm}t+\kappa^{(4)}_{nm},& \alpha_{19}=0,
\end{array}
\right.
$$
where
$$
\begin{array}{rcl}
A^{(i)}_{nm}&:=&\alpha_{i1}+\alpha_{i3}\kappa_{nm}^{(1)}+\alpha_{i5}\kappa_{nm}^{(2)}+\alpha_{i7}\kappa_{nm}^{(3)}, \\*[2ex]
B^{(i)}_{nm}&:=&\alpha_{i2}+\alpha_{i4}\kappa_{nm}^{(1)}+\alpha_{i6}\kappa_{nm}^{(2)}+\alpha_{i8}\kappa_{nm}^{(3)},
\end{array}
$$
for $i=1,2,3$ (functions with $i=2,3$ will also appear in what follows) and $\kappa^{(4)}_{nm}$ is an arbitrary function of $n,m$. Considering now the commutation relation  $[\widehat X_2,\widehat Y_9]$ we obtain
$$
\alpha_{19}\tau_0=0,\quad\quad\quad (\alpha_{29}+1)\tau_0=0,\quad\quad\quad \alpha_{29}a_{nm}=0,
$$
and
$$
\begin{array}{ll}
\left\{
\begin{array}{l}
\kappa^{(4)}_{nm}=0 \\
(2\alpha_{29}-1)B^{(1)}_{nm}-2\alpha_{19}B^{(2)}_{nm}=0 \\
(1+2\alpha_{29})\alpha_{19}A^{(1)}_{nm}+(1+2\alpha_{29})B^{(1)}_{nm}+2\alpha_{19}^2A^{(2)}_{nm}=0
\end{array}
\right. & \mbox{ for }\alpha_{19}\neq 0, \\*[2ex] \\
\left\{
\begin{array}{l}
(\frac{3}{2}-\alpha_{29})(B^{(1)}_{nm})^2=0 \\
(\alpha_{29+\frac{1}{2}})\kappa^{(4)}_{nm}+A^{(2)}_{nm}=0 \\
(\alpha_{29}-\frac{1}{2})A^{(1)}_{nm}+B^{(2)}_{nm}=0.
\end{array}
\right. & \mbox{ for }\alpha_{19}=0.
\end{array}
$$
From the commutation relation $[\widehat X_3,\widehat Y_9]$ we find
$$
\tau_0=0,\quad\quad\quad \alpha_{39}a_{nm}=0
$$
and
$$
\begin{array}{ll}
\left\{
\begin{array}{l}
\alpha_{19}\alpha_{39}A^{(1)}_{nm}-\alpha_{19}^2A^{(3)}_{nm}+\alpha_{39}B^{(1)}_{nm}=0\\
(1+\alpha_{19}\alpha_{39})B^{(1)}_{nm}-\alpha_{19}^2B^{(3)}_{nm}+\alpha_{19}A^{(1)}_{nm}=0
\end{array}
\right. & \mbox{ for }\alpha_{19}\neq 0, \\*[2ex] \\
\left\{
\begin{array}{l}
B^{(1)}_{nm}=0 \\
\kappa^{(4)}_{nm}+B^{(3)}_{nm}+\alpha_{39}A^{(1)}_{nm}=0\\
A^{(3)}_{nm}+\alpha_{39}\kappa^{(4)}_{nm}=0
\end{array}
\right. & \mbox{ for }\alpha_{19}=0.
\end{array}
$$

Finally, for $j=1,\ldots,8$ the commutation relations $[\widehat Y_j,\widehat Y_9]$ imply $\beta_{j9}a_{nm}=0$ and
$$
\begin{array}{ll}
\left\{
\begin{array}{l}
\kappa^{(\frac{j-1}{2})}a_{nm}=C_{nm}^{(j)}+\beta_{j9}\kappa^{(4)}_{nm}\\
D_{nm}^{(j)}+\beta_{j9}A^{(1)}_{nm}=0
\end{array}
\right. & \mbox{ for }j=1,3,5,7 \\*[2ex] \\
\left\{
\begin{array}{l}
\kappa^{(\frac{j-2}{2})}a_{nm}=D_{nm}^{(j)}+\beta_{j9}A^{(1)}_{nm}\\
C_{nm}^{(j)}+\beta_{j9}\kappa^{(4)}_{nm}=0
\end{array}
\right. & \mbox{ for }j=2,4,6,8
\end{array}
$$
where
$$
\begin{array}{rcl}
C^{(j)}_{nm}&:=&\beta_{j1}+\beta_{j3}\kappa_{nm}^{(1)}+\beta_{j5}\kappa_{nm}^{(2)}+\beta_{j7}\kappa_{nm}^{(3)}, \\*[2ex]
D^{(j)}_{nm}&:=&\beta_{j2}+\beta_{j4}\kappa_{nm}^{(1)}+\beta_{j6}\kappa_{nm}^{(2)}+\beta_{j8}\kappa_{nm}^{(3)},
\end{array}
$$
and $\kappa_{nm}^{(0)}:=1$ for convenience.

Since $A^{(i)}_{nm}$, $B^{(i)}_{nm}$, $C^{(j)}_{nm}$ and $D^{(j)}_{nm}$  are linear combinations of linearly independent functions appearing in the vector fields $\widehat Y_j$, we can use linear combinations to simplify $\widehat Y_9$. In the generic case, i.e. for any $\alpha_{19}$, we have that $a_{nm}=C_{nm}^{(1)}$ and using linear combinations we can transform $a_{nm}$ to zero. Moreover, we have that $\tau=0$ such that $\widehat Y_9=\lambda_{nm}(t)\partial_{u_{nm}}$, where $\lambda_{nm}$ depends on $\alpha_{19}$. In the case $\alpha_{19}\neq 0$ we can transform $A^{(1)}_{nm}=B^{(1)}_{nm}=0$ by linear combinations. From equations obtained previously, this implies that $A^{(i)}_{nm}=B^{(i)}_{nm}=0$ for $i=1,2,3$. Since $\kappa^{(4)}_{nm}=0$, no additional symmetry is possible when $\alpha_{19}\neq 0$. When $\alpha_{19}=0$, one equation gives us that $B^{(1)}_{nm}=0$ and by linear combinations we can transform $A^{(1)}_{nm}$ to zero. Therefore, we find
$$
\widehat Y_9=\kappa^{(4)}_{nm}\partial_{u_{nm}}.
$$
Note that we cannot use allowed transformations to simplify $\widehat Y_9$. All allowed transformations have been already used to simplify vector fields, in particular to obtain $\kappa_{nm}^{(0)}=1$ in $\widehat Y_1$ and $\widehat Y_2$.

From the remaining equation (\ref{remaining}) we have
$$
\begin{array}{rcl}
0&=& \displaystyle \sum_{(p,q)\in \Gamma} \kappa^{(4)}_{pq}\partial_{u_{pq}}F_{nm},
\end{array}
$$
where $F_{nm}$ is given by (\ref{functionF11}). Using the method of characteristics we find that the new invariant function when $\widehat Y_9$ is added to $NS_{11,1}$ is given by (\ref{functionF13}).

Let us now verify that the Lie algebra $\{\widehat X_1,\widehat X_2,\widehat X_3,\widehat Y_1,\ldots,\widehat Y_9\}$ is `maximal' or not, i.e. if we can add another vector field of the form $\widehat Y_{10}=\tau(t)\partial_t+\left[\left(\frac{\dot{\tau}}{2}+a_{nm}\right)u_{nm}+\lambda_{nm}(t)\right]\partial_{u_{nm}}$ with the same invariant function (\ref{functionF13}). From the Levi theorem, we have
$$
[\widehat X_i,\widehat Y_{10}]=\sum_{k=1}^{10} \widetilde \alpha_{ik} \widehat Y_k,\quad i=1,2,3 \quad\quad \mbox{ and } \quad\quad [\widehat Y_j,\widehat Y_{10}]=\sum_{k=1}^{10} \widetilde \beta_{jk} \widehat Y_k, \quad j=1,\ldots,9
$$
where $\widetilde \alpha_{ik},\widetilde \beta_{jk}$ are real constants. Now defining
$$
\begin{array}{rcl}
\widetilde A^{(i)}_{nm}&:=&\widetilde \alpha_{i1}+\widetilde \alpha_{i3}\kappa_{nm}^{(1)}+\widetilde \alpha_{i5}\kappa_{nm}^{(2)}+\widetilde \alpha_{i7}\kappa_{nm}^{(3)}+\widetilde \alpha_{i9}\kappa_{nm}^{(4)}, \\*[2ex]
\widetilde B^{(i)}_{nm}&:=&\widetilde \alpha_{i2}+\widetilde \alpha_{i4}\kappa_{nm}^{(1)}+\widetilde \alpha_{i6}\kappa_{nm}^{(2)}+\widetilde \alpha_{i8}\kappa_{nm}^{(3)}, \\*[2ex]
\widetilde C^{(j)}_{nm}&:=&\widetilde \beta_{j1}+\widetilde \beta_{j3}\kappa_{nm}^{(1)}+\widetilde \beta_{j5}\kappa_{nm}^{(2)}+\widetilde \beta_{j7}\kappa_{nm}^{(3)}+\widetilde \beta_{j9}\kappa_{nm}^{(4)}, \\*[2ex]
\widetilde D^{(j)}_{nm}&:=&\widetilde \beta_{j2}+\widetilde \beta_{j4}\kappa_{nm}^{(1)}+\widetilde \beta_{j6}\kappa_{nm}^{(2)}+\widetilde \beta_{j8}\kappa_{nm}^{(3)},
\end{array}
$$
the commutation relations $[\widehat X_i,\widehat Y_{10}]$ and $[\widehat Y_j,\widehat Y_{10}]$ give us the same set of equations obtain previously  replacing $A^{(i)}_{nm}\rightarrow \widetilde A^{(i)}_{nm},\ldots,D^{(j)}_{nm}\rightarrow \widetilde D^{(j)}_{nm},\alpha_{ik}\rightarrow \widetilde \alpha_{ik},\beta_{jk}\rightarrow \widetilde \beta_{jk}$ and $\kappa_{nm}^{(4)}\rightarrow \widetilde \kappa_{nm}^{(4)}$ for $i=1,2,3$, $j=1,\ldots,9$ and $k=1,\ldots, 10$. Again here we have that $\tau=0$ and $a_{nm}=\widetilde C_{nm}^{(1)}$. Using linear combinations we can transform $a_{nm}$ to zero such that $\widehat Y_{10}=\lambda_{nm}(t)\partial_{u_{nm}}$. For $\alpha_{1\,10}=0$ we can transform $\lambda_{nm}$ to zero by linear combination. In the case $\alpha_{1\,10}\neq 0$ we can transform $\widetilde A^{(1)}_{nm}$ to zero by linear combinations but not $\widetilde B^{(1)}_{nm}$ (since this function does not depend on $\kappa_{nm}^{(4)}$). Therefore, we obtain
$$
\widehat Y_{10}=\kappa^{(4)}_{nm}t\partial_{u_{nm}}.
$$
The remaining equation (\ref{remaining})
$$
\begin{array}{rcl}
0&=& \displaystyle \sum_{(p,q)\in \Gamma} \kappa^{(4)}_{pq}t\partial_{u_{pq}}F_{nm},
\end{array}
$$
with $F_{nm}$ given by (\ref{functionF13}), is identically zero. This complete the proof for the nonsolvable case of theorem~\ref{NonSolvable13}. $\Box$

\subsection*{Acknowledgements}
S.T. would like to thank Pavel Winternitz for interesting
discussions. The research of I.S.M. was partly supported by an
Institut des Sciences Math\'{e}matiques Scholarship. The research
of S.T. is partly supported by grant from NSERC of Canada.

\end{document}